\documentclass[preprint,aps,prd,floatfix,nofootinbib,10pt]{revtex4-2}
\pdfoutput=1
\usepackage{graphicx}
\usepackage{amsfonts}
\usepackage{nicefrac}
\usepackage{xcolor}
\usepackage{natbib}
\usepackage{braket}
\usepackage{hyperref}

\usepackage{amsmath}
\usepackage{rotating}
\usepackage{amssymb,amsthm}
\usepackage{soul}

\usepackage{xcolor}
\usepackage[normalem]{ulem}

\usepackage{latexsym}
\usepackage{graphics}

\usepackage{amsfonts}
\usepackage{amsmath}
\usepackage{rotating}
\usepackage{amssymb}

\usepackage{amsmath}
\usepackage{rotating}
\usepackage{amssymb}
\usepackage{soul}

\usepackage{xcolor}

\usepackage{xcolor}
\def\beq{\begin{eqnarray}}  
	\def\eeq{\end{eqnarray}}

\def \bh {\mbox{{\bf h}}}

\def\beq{\begin{eqnarray}}  
	\def\eeq{\end{eqnarray}}

\begin{document}
	
	\title{Chaplygin and polytropic gases Teleparallel Robertson-Walker $F(T)$ gravity solutions}
	
	\author{A. Landry}
	\email{a.landry@dal.ca}
	\affiliation{Department of Mathematics and Statistics, Dalhousie University, Halifax, Nova Scotia, Canada, B3H 3J5}

\begin{abstract}
	
	This paper investigates the teleparallel Robertson--Walker (TRW) $F(T)$ gravity solutions for a Chaplygin gas, and then for any polytropic gas cosmological source. We use the TRW $F(T)$ gravity field equations (FEs) for each $k$-parameter value case and the relevant gas equation of state (EoS) to find the new teleparallel $F(T)$ solutions. For flat $k=0$ cosmological case, we find analytical solutions valid for any cosmological scale factor. For curved $k=\pm 1$ cosmological cases, we find new approximated teleparallel $F(T)$ solutions for slow, linear, fast and very fast universe expansion cases summarizing by a double power-law function. All the new solutions will be relevant for future cosmological applications on dark matter, dark energy (DE) quintessence, phantom energy, Anti-deSitter (AdS) spacetimes and several other cosmological processes.

\end{abstract}

\maketitle
\tableofcontents

\newpage

	\section{Introduction}\label{sec1}

The teleparallel $F(T)$ gravity is a frame-based and alternative theory to general relativity (GR) fundamentally defined in terms of the {coframe} ${\bf h}^a$ and the spin-connection $\omega^a_{~bc}$~\cite{Lucas_Obukhov_Pereira2009,Aldrovandi_Pereira2013,Bahamonde:2021gfp,Krssak:2018ywd,MCH,Coley:2019zld,Krssak_Pereira2015}. These two last quantities define the torsion tensor $T^a_{~bc}$ and torsion scalar $T$. {We remind that GR is defined by the metric $g_{\mu\nu}$ and the spacetime curvatures $R^a_{~b\mu\nu}$, $R_{\mu\nu}$ and $R$.} Under some considerations, we can determine the symmetries for any independent coframe/spin-connection pairs, and then spacetime curvature and torsion are defined as geometric objects~\cite{MCH,Coley:2019zld,Krssak_Saridakis2015,Krssak_Pereira2015,Aldrovandi_Pereira2013,olver1995equivalence}. Any geometry described by a such pair whose curvature and non-metricity are both zero ($R^a_{~b\mu\nu}=0$ and $Q_{a\mu\nu}=0$ conditions) is a teleparallel gauge-invariant geometry (valid for any $g_{ab}$). The fundamental pairs must satisfy two Lie derivative-based relations and we use the Cartan--Karlhede algorithm to solve these two fundamental equations for any teleparallel geometry. For a pure teleparallel $F(T)$ gravity spin-connection solution, we also solve the null Riemann curvature condition leading to a Lorentz transformation-based definition of the spin-connection $\omega^a_{~b\mu}$. There is a direct equivalent to GR in teleparallel gravity: the teleparallel equivalent to GR (TEGR) which generalizes to teleparallel $F(T)$-type gravity \cite{Aldrovandi_Pereira2013,Krssak_Saridakis2015,Ferraro:2006jd,Ferraro:2008ey,Linder:2010py}. All the previous considerations are also adapted for the new general relativity (NGR) (refs.~\cite{kayashi,beltranngr,bahamondengr} and refs. therein), the symmetric teleparallel $F(Q)$-type gravity (refs.~\cite{heisenberg1,heisenberg2,faithman1,hohmannfq} and refs. therein) and some extended theories like $F(T,Q)$-type, $F(R,Q)$-type, $F(R,T)$-type, and several other ones (refs.~\cite{jimeneztrinity,nakayama,ftqgravity,frtspecial,frttheory,myrzakulov1,myrzakulov2,myrzakulov3,myrzakulov4,myrzakulov5} and refs. therein). In the current paper we will restrict our study to the teleparallel $F(T)$ gravity framework.

There is a large number of research papers on spherically symmetric spacetimes and solutions in teleparallel $F(T)$ gravity using a large number of approaches, energy-momentum sources and made for various purposes~\cite{golov1,golov2,golov3,debenedictis,SSpaper,TdSpaper,nonvacSSpaper,nonvacKSpaper,staticscalarfieldSS,scalarfieldKS,roberthudsonSSpaper,coleylandrygholami,scalarfieldTRW,baha1,bahagolov1,awad1,baha6,nashed5,pfeifer2,elhanafy1,benedictis3,baha10,baha4,ruggiero2,sahoo1,sahoo2,calza}. But there is a special class of teleparallel spacetime in which the field equations (FEs) are purely symmetric: the teleparallel Robertson--Walker (TRW) spacetime \cite{preprint,coleylandrygholami,scalarfieldTRW}. This type of spacetime is defined in terms of the $k$-parameter where $k=0$ is a flat cosmological spacetime, $k=\pm 1$ cases are, respectively, positive and negative space cosmological curvature \cite{aldrovandi2003,bounce,Capozz,inflat}. We have proven that a such teleparallel geometry is described by a $G_6$ Lie algebra group, where the fourth to sixth Killing Vectors (KVs) are characteristic of this spacetime \cite{preprint,coleylandrygholami}. The main apparent consequence of additional KVs is the trivial antisymmetric parts of FEs. The same spacetime geometry structure also exists for some extensions such as teleparallel $F(T,B)$ gravity \cite{FTBcosmogholamilandry,HJKP2018,Cai_2015,dixit,ftbcosmo3}. We have found teleparallel $F(T)$ and $F(T,B)$ solutions for perfect fluid (PF) and scalar field (SF) sources. In the last case, the teleparallel $F(T)$ and $F(T,B)$ solutions are scalar potential-independent and only SF-dependent \cite{scalarfieldTRW,FTBcosmogholamilandry}. But there are additional possible sources of energy-momentum which can lead to new further teleparallel solutions in $F(T)$-type, but also for extensions.

The dark energy (DE) states are usually studied by using the PF equivalent equation of state (EoS) $P_{\phi}=\alpha_Q\,\rho_{\phi}$, where $\alpha_Q$ is the DE index (or quintessence index in some refs.) \cite{hawkingellis1,cosmofluidsbohmer,BahamondeBohmer}. The possible DE forms in terms of $\alpha_Q$ are as follows:
\begin{enumerate}
	\item {\textbf{Quintessence} ${\bf -1<}\boldmath{\alpha_Q} {\bf <-\frac{1}{3}}$:} 
 This describes a controlled accelerating universe expansion where energy conditions are always satisfied, i.e., $P_{\phi}+\rho_{\phi} > 0$ \cite{steinhardt1,steinhardt2,steinhardt3,carroll1,quintessencecmbpeak,quintessenceholo,quintchakra2024,cosmofate,rollingscalarfield,steinhardt2024,wolf1,wolf2,wolf3}. This usual DE form has been significantly studied in the literature in recent decades for the fascination it provokes and the realism of the models.
	
	\item {\textbf{Phantom energy} $\boldmath{\alpha_Q} {\bf <-1}$}: This describes an uncontrolled universe expansion accelerating toward a Big Rip event { (or singularity)} \cite{quintessencephantom,strongnegative,farnes,baumframpton,caldwell1,phantomdivide,phantomteleparallel1,ripphantomteleparallel2,phantomteleparallel3}. The energy condition is violated, i.e., $P_{\phi}+\rho_{\phi} \ngeq 0$. { But this DE form is fascinating because we can find new teleparallel solutions and physical models.}
	
	\item {\textbf{Cosmological constant} $\boldmath{\alpha_Q}{\bf =-1}$}: This is an intermediate limit between the quintessence and phantom DE states, where $P_{\phi}+\rho_{\phi} = 0$. A constant SF source $\phi=\phi_0$ added by a positive scalar potential $V\left(\phi_0\right)>0$ will directly lead to this primary DE state. Note that a negative scalar potential (i.e., $V\left(\phi_0\right)\leq 0$) will not lead to a positive cosmological constant and/or a DE solution.	
	
	\item {\textbf{Quintom models}}: This is a mixture of previous DE types, usually described by double SF models \cite{quintom1,quintom2,quintom3,quintom4,quintomcoleytot,quintomteleparallel1}. This type of model is more complex to study and solve in general. Several types of models are in principle possible, and these physical processes need further studies in the future.
\end{enumerate}
There are, in the recent literature on teleparallel gravity and its extension, a number of papers discussing possible solutions in static, radial-dependent, time-dependent and cosmological teleparallel $F(T)$ models \cite{nonvacSSpaper,staticscalarfieldSS,nonvacKSpaper,scalarfieldKS,scalarfieldTRW,coleylandrygholami,FTBcosmogholamilandry,roberthudsonSSpaper,myrzacosmotele1,myrzacosmospin,paliacosmo}. Most of those are suitable for DE models, in particular for quintessence DE state. We use cosmological PF as SF approaches, and we find that the solution classes are similar in the both cases. { Some of these papers aimed to test some simple teleparallel $F(T)$ solutions, such as power-law and the polynomial solution, on the cosmological models such as the well-known Lambda Cold Dark Matter ($\Lambda$CDM) model limits \cite{myrzacosmotele1,myrzacosmospin,paliacosmo}. Therefore, we also need to test some non-linear PF-based universe models and determine the non-linearity level of a typical cosmological PF-based universe.}

The quartessence model is a unified model of DE and Dark Matter (DM) arising from a Chaplygin cosmological gas. The DE and DM are two states of a single and simple quartessence dark cosmological fluid \cite{Kamenshchik2001,Bento2002,Bilic:2002chg,zhu2004,Makler:2003iw}. This non-linear cosmological fluid model can be considered as an alternative and unified explanation of DE and DM evolving in the universe and influencing the cosmological processes. For polytropic gases, there are additional possible physical models arising from this class of EoS-based cosmological solutions \cite{polytropic1,polytropic2,polytropic3,polytropic4,polytropic5,polytropic6,polytropic7,polytropic8}. But the Chaplygin, polytropic gas and any superposition of those models in general can explain not only DE and DM mixed or separate models, but also any non-linear cosmological gas and fluid system at the limit \cite{Arun:2017dm}. Most of the previous works have been done in the GR (or $F(R)$-type) framework, but these last cases may also arise in teleparallel theories of gravity. Under this last consideration, the current investigation concerning the Chaplygin and polytropic fluids deserves to be achieved in  teleparallel framework and constitutes the main aim of the paper, in particular for the $F(T)$-type case. Ultimately, we want to build a pure teleparallel quartessence and polytropic-based cosmology, allowing us to study more realistic universe models.

Ultimately we want to study in detail the quartessence suitable teleparallel cosmological solutions and its physical impacts. Therefore we need, at the current stage, to find the possible Chaplygin and polytropic cosmological gas teleparallel $F(T)$ gravity solutions in a Robertson--Walker spacetime (TRW). We had found the TRW geometry and we had solved the TRW FEs and conservation laws (CL) for PF and SF solutions in teleparallel $F(T)$ and $F(T,B)$ gravities \cite{coleylandrygholami,preprint,FTBcosmogholamilandry,scalarfieldTRW}. But we can go further and aim to solve for polytropic and Chaplygin cosmological gases teleparallel $F(T)$ solutions as the next step. To satisfy this aim, we will use the same TRW geometry, FEs and CLs to develop the teleparallel Chaplygin $F(T)$ solutions in Section \ref{sect3} and the teleparallel polytropic $F(T)$ solutions in Section \ref{sect4}. We will then compare and highlight the similarities and differences by using graphs { and propose future experimental data-based analysis guidelines} in Section~\ref{sect5} before concluding the current paper in Section \ref{sectconcl}.


\section{Summary of Teleparallel Gravity and Field Equations}\label{sect2}

\subsection{Teleparallel $F(T)$-Gravity Theory Field Equations and Torsional Quantities}\label{sect21}

The teleparallel $F(T)$-type gravity action integral with any gravitational source is \cite{Aldrovandi_Pereira2013,Bahamonde:2021gfp,Krssak:2018ywd,Coley:2019zld,SSpaper,nonvacSSpaper,nonvacKSpaper,roberthudsonSSpaper,scalarfieldKS,scalarfieldTRW,staticscalarfieldSS}:
\begin{equation}\label{1000}
	S_{F(T)} = \int\,d^4\,x\,\left[\frac{h}{2\kappa}\,F(T)+\mathcal{L}_{Source}\right],
\end{equation}
where $h$ is the coframe determinant, $\kappa$ is the coupling constant and $\mathcal{L}_{Source}$ is the gravitational source term. We will apply the least-action principle on Equation \eqref{1000} to find the symmetric and antisymmetric parts of FEs as \cite{SSpaper,nonvacSSpaper,nonvacKSpaper,roberthudsonSSpaper,scalarfieldKS,scalarfieldTRW,staticscalarfieldSS}:
\begin{eqnarray}
	\kappa\,\Theta_{\left(ab\right)} &=& F_T\left(T\right) \overset{\ \circ}{G}_{ab}+F_{TT}\left(T\right)\,S_{\left(ab\right)}^{\;\;\;\mu}\,\partial_{\mu} T+\frac{g_{ab}}{2}\,\left[F\left(T\right)-T\,F_T\left(T\right)\right],  \label{1001a}
	\\
	0 &=& F_{TT}\left(T\right)\,S_{\left[ab\right]}^{\;\;\;\mu}\,\partial_{\mu} T, \label{1001b}
\end{eqnarray}
with $\overset{\ \circ}{G}_{ab}$ the Einstein tensor, $\Theta_{\left(ab\right)}$ the energy-momentum, $g_{ab}$ the gauge metric and $\kappa$ the coupling constant. The torsion tensor $T^a_{~~\mu\nu}$, the torsion scalar $T$ and the super-potential $S_a^{~~\mu\nu}$ are defined as \cite{Coley:2019zld}
\begin{align}
	T^a_{~~\mu\nu} =& \partial_{\mu}\,h^a_{~~\nu}-\partial_{\nu}\,h^a_{~~\mu}+\omega^a_{~~b\mu}h^b_{~~\nu}-\omega^a_{~~b\nu}h^b_{~~\mu}, \label{torsionten}
	\\
	S_a^{~~\mu\nu}=& \frac{1}{2}\,\left(T_a^{~~\mu\nu}+T^{\nu\mu}_{~~a}-T^{\mu\nu}_{~~a}\right)-h_a^{~~\nu}\,T^{\lambda\mu}_{~~\lambda}+h_a^{~~\mu}\,T^{\lambda\nu}_{~~\lambda},
	\\
	T=&\frac{1}{2}\,T^a_{~~\mu\nu}S_a^{~~\mu\nu}.
\end{align}	
{Equation} 
 \eqref{torsionten} can be expressed in terms of the three irreducible parts of torsion tensor as
\begin{align}
	T_{abc} = \frac{2}{3}\left(t_{abc}-t_{acb}\right)-\frac{1}{3}\left(g_{ab}V_c-g_{ac}V_b\right)+\epsilon_{abcd}A^d
\end{align}
where,
\begin{align}
	V_a=&T^b_{~ba} ,
	\quad
	A^a=\frac{1}{6}\epsilon^{abcd}T_{bcd} ,
	\quad
	t_{abc}= \frac{1}{2}\left(T_{abc}+T_{bac}\right)-\frac{1}{6}\left(g_{ca}V_b+g_{cb}V_a\right)+\frac{1}{3}V_c.
\end{align}
{We} usually solve in teleparallel $F(T)$ gravity, Equations \eqref{1001a} and \eqref{1001b}. Therefore in refs. \cite{preprint,coleylandrygholami,scalarfieldTRW}, we showed that Equation \eqref{1001b} is trivially satisfied despite a non-zero spin-connection, because the teleparallel geometry is purely symmetric. Only Equation \eqref{1001a} is non-trivial and will be explicitly solved in detail.

\subsection{Teleparallel Robertson--Walker Spacetime Geometry}\label{sect22}

Any frame-based geometry in teleparallel gravity on a frame bundle is defined by a coframe/spin-connection pair and a field ${\bf X}$. The geometry must satisfy the fundamental Lie derivative-based equations \cite{Coley:2019zld,MCH,preprint,coleylandrygholami,scalarfieldTRW}:
\begin{equation}
	\mathcal{L}_{{\bf X}}\,\bh_a = \lambda_a^{~b} \,\bh_b \mbox{ and } \mathcal{L}_{{\bf X}}\, \omega^a_{~bc} = 0, \label{Intro:FS2}
\end{equation}
where $\omega^a_{~bc}$ is the spin-connection in terms of the differential coframe $\bh_a$ and $\lambda_a^{~b}$ is the linear isotropy group component. In addition for a pure teleparallel $F(T)$-type gravity, we must also satisfy the null Riemann curvature condition $R^a_{~abc}=0$. For TRW spacetime geometries on an orthonormal frame, the coframe/spin-connection pair  $h^a_{\;\;\mu}$ and $\omega_{abc}$ solutions are \cite{preprint,coleylandrygholami,scalarfieldTRW} 
\begin{align}\label{TRWcoframe}
	h^a_{\;\;\mu} =& Diag\left[1, a(t)\,\left(1-k\,r^2\right)^{-1/2},\,a(t)\,r,\, a(t)\,r\,\sin\theta\right],
	\\
	\omega_{122} =& \omega_{133} = \omega_{144} =  W_1(t), \quad  \omega_{234} = -\omega_{243} = \omega_{342} = W_2(t), \quad
	\nonumber\\
	 \omega_{233} =& \omega_{244} = - \frac{\sqrt{1-kr^2}}{a(t)r}, \quad
	\omega_{344} =  \frac{\cot(\theta)}{a(t) r}, \label{Con:FLRW} 
\end{align}
where $W_1$ and $W_2$ are depending on $k$-parameter and defined by
\begin{enumerate}
	\item $k=0$: $W_1=W_2=0$,
	\item $k=+1$: $W_1=0$ and $W_2(t)=\pm\,\frac{\sqrt{k}}{a(t)}$,	
	\item $k=-1$: $W_1(t)=\pm\,\frac{\sqrt{-k}}{a(t)}$ and $W_2=0$.		
\end{enumerate}
{For} any $W_1$ and $W_2$, we will obtain the same symmetric FEs set to solve for each subcases depending on $k$-parameter. The previous coframe/spin-connection pair was found by solving  Equation \eqref{Intro:FS2} and $R^a_{~b\mu\nu}=0$ condition as defined in ref. \cite{Coley:2019zld}. These solutions were also used in $F(T)$ TRW spacetime recent works \cite{preprint,coleylandrygholami,scalarfieldTRW}. The TRW spacetime structure is typically explanable by a $G_6$ Lie algebra group. The FEs to be solved in the current paper are defined for each $k$-parameter cases and will lead to additional new teleparallel $F(T)$ solution classes. The FEs defined by Equations \eqref{1001a} and \eqref{1001b} are still purely symmetric and valid on proper frames, as showed in refs. \cite{preprint,coleylandrygholami,scalarfieldTRW}. Equation \eqref{1001b} are trivially satisfied, and we will solve Equation \eqref{1001a} for each $k$-parameter case.

\subsection{Conservation Laws and Field Equations of Cosmological Perfect Fluids}\label{sect23}

The canonical energy-momentum and its GR CLs are obtained from $\mathcal{L}_{Source}$ term of Equation \eqref{1000} as \cite{Aldrovandi_Pereira2013,Bahamonde:2021gfp}
\begin{align}
	\Theta_a^{\;\;\mu}=&\frac{1}{h} \frac{\delta \mathcal{L}_{Source}}{\delta h^a_{\;\;\mu}}, \quad
	\Rightarrow\quad \overset{\ \circ}{\nabla}_{\nu}\left(\Theta^{\mu\nu}\right)=0 , \label{1001e}
\end{align}
where $\overset{\ \circ}{\nabla}_{\nu}$ the covariant derivative and $\Theta^{\mu\nu}$ the conserved energy-momentum tensor. The antisymmetric and symmetric parts of $\Theta_{ab}$ are \cite{SSpaper,nonvacSSpaper,nonvacKSpaper,roberthudsonSSpaper,scalarfieldKS,scalarfieldTRW,staticscalarfieldSS}
\begin{equation}\label{1001c}
	\Theta_{[ab]}=0,\qquad \Theta_{(ab)}= T_{ab},
\end{equation}
where $T_{ab}$ is the symmetric part of $\Theta^{\mu\nu}$. Equation \eqref{1001e} also imposes the symmetry of $\Theta^{\mu\nu}$ and Equation \eqref{1001c} the condition. Equation \eqref{1001c} is valid only when the matter field interacts with the metric $g_{\mu\nu}$ defined from the coframe $h^a_{\;\;\mu}$ and the gauge $g_{ab}$, and is not directly coupled to the $F(T)$ gravity. This consideration is only valid for the null hypermomentum case (i.e., $\mathfrak{T}^{\mu\nu}=0$) as discussed in refs. \cite{golov3,nonvacSSpaper,nonvacKSpaper,roberthudsonSSpaper,scalarfieldKS,scalarfieldTRW,staticscalarfieldSS}. This last condition on hypermomentum is defined from Equations \eqref{1001a} and \eqref{1001b} as \cite{golov3}
\begin{align}\label{1001h}
	\mathfrak{T}_{ab}=\kappa\Theta_{ab}-F_T\left(T\right) \overset{\ \circ}{G}_{ab}-F_{TT}\left(T\right)\,S_{ab}^{\;\;\;\mu}\,\partial_{\mu} T-\frac{g_{ab}}{2}\,\left[F\left(T\right)-T\,F_T\left(T\right)\right]=0.
\end{align}
There are more general teleparallel $\mathfrak{T}^{\mu\nu}$ definitions and $\mathfrak{T}^{\mu\nu}\neq 0$ CLs, but this does not really concern the teleparallel $F(T)$-gravity situation \cite{hypermomentum1,hypermomentum2,hypermomentum3,golov3}.

For a TRW spacetime geometry defined by Equations \eqref{TRWcoframe} and \eqref{Con:FLRW}, Equation \eqref{1001e} for a $P=P(\rho)$ fluid is \cite{preprint,coleylandrygholami,scalarfieldTRW}
\begin{align}\label{1003}
	\dot{\rho}+3\,H\,\left(P(\rho)+\rho\right)=0,
\end{align}
where $H=\frac{\dot{a}}{a}$ is the Hubble parameter. The general FEs system for TRW cosmological spacetimes are \cite{preprint,coleylandrygholami,scalarfieldTRW}
\begin{enumerate}
	\item {${\bf k=0}$ \textbf{flat or non-curved}}:
	\begin{align}
		\kappa\rho	= &	-\frac{F}{2}+6H^2\,F_T ,  \label{1010}
		\\
		\kappa(\rho+3P(\rho)) =&	F-6\left( \dot{H}+2H^2\right)F_T-6H\,F_{TT}\dot{T}, \label{1011}
		\\
		T =& 6H^2  .\label{1012}
	\end{align}
	{The} Equation \eqref{1012} yields to $H=\sqrt{\frac{T}{6}}$, and then Equations \eqref{1010} and \eqref{1011} become
	\begin{align}
		\kappa\rho	= &	-\frac{F}{2}+T\,F_T ,  \label{1013}
		\\
		\kappa(\rho+3P(\rho)) =&	F-\left( 6\dot{H}+2T\right)F_T-12\dot{H}\,T\,F_{TT}, \label{1014}
	\end{align}
	{The} pure vacuum solution ($\rho=0$ and $P=0$) to Equations \eqref{1013} and \eqref{1014} is $F(T)=F_0\,\sqrt{T}$. However for any $P=P(\rho)$, we can set $a(t)=a_0\,t^n$ as cosmological scale and $F(T)=\sqrt{T}\,G(T)$ as solution ansatz, and we find the unified FE by merging Equations \eqref{1013} and \eqref{1014}:
	\begin{align}
		\frac{3}{2}\left(n-1+\frac{n\,P(\rho)}{\rho}\right)G_T =&  T\,G_{TT} . \label{1015}
	\end{align}
	Equation \eqref{1015} is the general $k=0$ unified FE to solve for any EoS and CL. This is easy to solve, and the solution will be an easy-to-compute integral equation.

	\item {${\bf k=-1}$ \textbf{negative curved}}:
	\begin{align}
		\kappa\rho	= & -\frac{F}{2}+6\,H\left(H+\frac{\delta\sqrt{-k}}{a}\right) F_T, \label{1020}
		\end{align}
		
		\begin{align}
\kappa(\rho+3P(\rho)) =&	F-6\left(\dot{H}+H^2+\left(H+\frac{\delta\sqrt{-k}}{a}\right)^2 \right) F_T -6\left(H+\frac{\delta\sqrt{-k}}{a}\right)\,F_{TT}\dot{T}, \label{1021}
		\end{align}
		\begin{align}
		T =& 6\left( H+ \frac{\delta\sqrt{-k}}{a}\right)^2  . \label{1022}
	\end{align}
	{From} Equation \eqref{1022} and using $a(t)=a_0\,t^n$ ansatz, we find a characteristic equation yielding to $t(T)$ solutions:
	\begin{align}
		& 0= \frac{\delta\sqrt{-k}}{a_0}\,t^{-n}+n t^{-1}-\delta_1\sqrt{\frac{T}{6}}. \label{1023}
	\end{align}
	{The} possible solutions of Equation \eqref{1023} are
	\begin{enumerate}
		\item {${\bf n=\frac{1}{2}}$} (slow expansion): 
		\begin{align}
			& t^{-1}(T)=\left[-\frac{\delta\sqrt{-k}}{a_0}\pm \sqrt{-\frac{k}{a_0^2}+\delta_1\sqrt{\frac{2T}{3}}}\right]^2. \label{1023a}
		\end{align}
		
		\item {${\bf n=1}$} (linear expansion):
		\begin{align}
			&  t^{-1}(T)=\frac{\delta_1}{\left(\frac{\delta\sqrt{-k}}{a_0}+1\right)}\sqrt{\frac{T}{6}}. \label{1023b}
		\end{align}
		
		\item {${\bf n=2}$} (fast expansion):
		\begin{align}
			& t^{-1}(T)=-\frac{\delta a_0}{\sqrt{-k}}\pm\sqrt{-\frac{a_0^2}{k}+\delta_1 \delta a_0\sqrt{-\frac{T}{6k}}}
			. \label{1023c}
		\end{align}

		\item {${\bf n\,\rightarrow\,\infty}$} (very fast expansion limit):	
		\begin{align}
			& t^{-1}(T)\approx \frac{\delta_1}{n}\sqrt{\frac{T}{6}}\,\rightarrow\,0. \label{1023d}
		\end{align}
	\end{enumerate}
	
	Then Equations \eqref{1020} and \eqref{1021} become by substituting Equation \eqref{1023} and the ansatz:
	\begin{align}
		\kappa\rho	= & -\frac{F}{2}+\sqrt{6}\,n\delta_1\,t^{-1}(T) \sqrt{T} F_T, \label{1024}
		\\
		\kappa\rho\left(1+\frac{3P(\rho)}{\rho}\right) =&	F-\left(6n(n-1)t^{-2}(T)+T \right) F_T 
		\nonumber\\
		&\;+12nt^{-1}(T)\,\left((1-n)\,t^{-1}(T)+\delta_1\sqrt{\frac{T}{6}}\right)T\,F_{TT}, \label{1025}
	\end{align}
	{The} unified FE from Equations \eqref{1024}and \eqref{1025} is
	
\begin{align}
		& \left[ -\frac{F}{2}+\sqrt{6}\,n\delta_1\,t^{-1}(T) \sqrt{T} F_T\right]\left(1+\frac{3P(\rho)}{\rho}\right) 
		\nonumber\\
		&\quad =	F-\left(6n(n-1)t^{-2}(T)+T \right) F_T +12nt^{-1}(T)\,\left((1-n)\,t^{-1}(T)+\delta_1\sqrt{\frac{T}{6}}\right)T\,F_{TT}, \label{1026}
	\end{align}

	\item {${\bf k=+1}$ \textbf{positive curved}:}
	\begin{align}
		\kappa\rho	= &	-\frac{F}{2}+6H^2\,F_T ,  \label{1030}
		\\
		\kappa(\rho+3P(\rho)) =&	F-6\left( \dot{H}+2H^2-\frac{k}{a^2}
		\right)F_T-6H\,F_{TT}\dot{T}, \label{1031}
		\\
		T =& 6\left[ H^2- \frac{k}{a^2}\right]  . \label{1032}
	\end{align}
	From Equation \eqref{1032} and using $a(t)=a_0\,t^n$ ansatz, we find the characteristic equation for $t^{-1}(T)$:
	\begin{align}
		&\frac{k}{a_0^2}t^{-2n}-n^2\,t^{-2}+\frac{T}{6}=0 . \label{1033}
	\end{align}
	{The} possible solutions of Equation \eqref{1033} are
	\begin{enumerate}
		\item {${\bf n=\frac{1}{2}}$} (slow expansion):
		\begin{align}
			&t^{-1}(T)= \frac{2k}{a_0^2}\pm\sqrt{\frac{4k^2}{a_0^4}+\frac{2T}{3}}  . \label{1033a}
		\end{align}
		
		\item {${\bf n=1}$} (linear expansion):
		\begin{align}
			t^{-2}(T)=\frac{T}{6\left(1-\frac{k}{a_0^2}\right)}. \label{1033b}
		\end{align}
		
		\item {${\bf n=2}$} (fast expansion):
		\begin{align}
			& t^{-2}(T)= \frac{2a_0^2}{k}\pm\sqrt{\frac{4a_0^4}{k^2}-\frac{a_0^2\,T}{6k}}	. \label{1033c}
		\end{align}
		
		\item ${\bf n\,\rightarrow\,\infty}$ (very fast expansion limit):	
		\begin{align}
			t^{-2}(T)\,\approx\frac{T}{6n^2}\,\rightarrow\,0. \label{1033d}
		\end{align}
	\end{enumerate}
	Then Equations \eqref{1030}--\eqref{1031} become by substituting Equation \eqref{1033} and the ansatz:
	\begin{align}
		\kappa\rho	= &	-\frac{F}{2}+6n^2\,t^{-2}(T)\,F_T ,  \label{1034}
		\\
		\kappa\rho\left(1+\frac{3P(\rho)}{\rho}\right) =&	F-\left( 6n(n-1)\,t^{-2}(T)+T
		\right)F_T
		\nonumber\\
		&\;+12n^2\,t^{-2}(T)\left(T-6n(n-1)t^{-2}(T)\right)\,F_{TT}, \label{1035}
	\end{align}
	The unified FE from Equations \eqref{1034}--\eqref{1035} is
	\begin{align}
		&\left[-\frac{F}{2}+6n^2\,t^{-2}(T)\,F_T \right]\left(1+\frac{3P(\rho)}{\rho}\right) 
		\nonumber\\
		&\quad =	F-\left( 6n(n-1)\,t^{-2}(T)+T
		\right)F_T+12n^2\,t^{-2}(T)\left(T-6n(n-1)t^{-2}(T)\right)\,F_{TT}, \label{1036}
	\end{align}

\end{enumerate}

\subsection{Energy Conditions in Teleparallel Gravity}\label{sect24}

Regardless of the definition of EoS or any pressure-density relationship, there are energy conditions to satisfy for any physical system based on a PF \cite{Kontou:2020bta}:
\begin{itemize}
	\item Weak Energy Condition ({\textbf{WEC}}): $\rho \geq 0$, $P_r+\rho \geq 0$ and $P_t+\rho \geq 0$.
	
	\item Strong Energy Condition ({\textbf{SEC}}): $P_r+2P_t+\rho \geq 0$, $P_r+\rho \geq 0$ and $P_t+\rho \geq 0$.

	\item Null Energy Condition ({\textbf{NEC}}): $P_r+\rho \geq 0$ and $P_t+\rho \geq 0$. 
	
	\item Dominant Energy Condition ({\textbf{DEC}}): $\rho \geq |P_r|$ and $\rho \geq |P_t|$.
\end{itemize}
{From} this consideration and by setting $P_r=P_t$ for a uniform pressure cosmological fluid, we can summarize the energy conditions (ECs):
\begin{align}
	& \rho+P \geq 0 , \quad\rho+3P\geq 0 , \quad \rho \geq |P|,   \quad \rho \geq 0.   \label{1040}
\end{align}
{In} Sections \ref{sect3} and \ref{sect4}, we will apply Equations \eqref{1040} for any CL solutions. In this way, we will verify the physical consistency of all CL solutions found in the current paper.

\section{Pure Chaplygin Gas Teleparallel Field Equation Solutions}\label{sect3}

\subsection{Conservation Law Solutions and Energy Conditions}\label{sect30}

For a pure Chaplygin gas of EoS $P=\frac{P_0}{\rho}$ (see refs. \cite{Kamenshchik2001,Bento2002,Bilic:2002chg,zhu2004,Makler:2003iw,Arun:2017dm}) and by using a power-law ansatz $a(t)=a_0\,t^n$, Equation \eqref{1003} becomes
\begin{align}\label{400}
	&\dot{\rho}+\frac{3n}{t}\,\left(\frac{P_0}{\rho}+\rho\right)=0,
	\nonumber\\
	&\Rightarrow\,\rho(t)=\left[\left(P_0+\rho_0^2\right)\,t^{-6n}-P_0\right]^{1/2} .
\end{align}
{We} will apply the ECs as stated by Equations \eqref{1040} on Equation \eqref{400} and find that
\begin{align}\label{401}
	& \left(P_0+\rho_0^2\right)\,t^{-6n}\geq 0 , \quad\left(P_0+\rho_0^2\right)\,t^{-6n}\geq -2P_0 , \quad \left(P_0+\rho_0^2\right)\,t^{-6n} \geq P_0+ |P_0|,   \quad
	\nonumber\\
	& \left(P_0+\rho_0^2\right)\,t^{-6n}\geq P_0. 
\end{align}
{The} dominating condition from Equations \eqref{401} is 
\begin{align}\label{401a}
	& \left(P_0+\rho_0^2\right)\,t^{-6n} \geq 2|P_0|. 
\end{align}
{Note} that the dust matter limit (i.e., $P(\rho)\,\rightarrow\,0$ for any $\rho$) is only possible under the $P_0\,\rightarrow\,0$ limit for a Chaplygin gas or any EoS in the $P=P_0\, \rho^{-q}$ form, where $q>0$. Under the $q\,\rightarrow\,\infty$ limit, we find that $P\,\rightarrow\,0$ for any value of $P_0$ in this such case, another way to the dust matter limit. The ECs, as defined by Equation \eqref{1040}, are more trivially satisfied under the ordinary matter limit.

\subsection{$k=0$ Cosmological Solutions}\label{sect31}

Using Equation \eqref{400} for CL and EoS and the torsion scalar defined by Equation \eqref{1012}, the unified FEs described by Equation \eqref{1015} and the solution are
\begin{align} 
	&	\frac{3}{2}\left[n-1+\frac{nP_0}{\left[\left(P_0+\rho_0^2\right)\,T^{3n}-P_0\right]}\right] G_T=\,  T\,G_{TT}  ,
	\nonumber\\
	& \Rightarrow\,G_T(T)=G_T(0)\,T^{\frac{3}{2}(n-1)}\sqrt{\beta T^{3n}-P_0} , \label{403}
\end{align}		
{where $\beta=\frac{P_0+\rho_0^2}{\left(6n^2\right)^{3n}}$. By integration and substitution into $F(T)$ ansatz, we find as final solution:}
\begin{align}
	F(T)=G(0)\sqrt{T}-2G_T(0)\,\sqrt{-P_0}\,_2F_1 \left(-\frac{1}{2},\,-\frac{1}{6n};\,1-\frac{1}{6n};\frac{\beta}{P_0}T^{3n}\right). \label{404}
\end{align}	
{{As} in refs \cite{coleylandrygholami,scalarfieldTRW}, the $k=0$ flat cosmological case yields  an easy-to-compute analytical teleparallel $F(T)$ solution for any value of $n$. The first term of Equation \eqref{404} is recurrent to any cosmological teleparallel solution and constitutes a ``teleparallel background solution'' \cite{myrzacosmotele1,myrzacosmospin,paliacosmo}. The second term will vanish under the $n\,\rightarrow\,\infty$ limit, leading to $F(T)=G(0)\sqrt{T}$  the teleparallel background solution. There is no direct TEGR limit according to Equations \eqref{403} and \eqref{404} forms.}

\subsection{$k=-1$ Cosmological Solutions}\label{sect32}

Equation \eqref{1026} with $P(\rho)=\frac{P_0}{\rho}$ is solved by substituting Equation \eqref{400} solution:
\begin{align}
	& \left[ -\frac{F}{2}+\sqrt{6}\,n\delta_1\,t^{-1}(T) \sqrt{T} F_T\right]\left(\frac{\left(P_0+\rho_0^2\right)\,t^{-6n}(T)+2P_0}{\left(P_0+\rho_0^2\right)\,t^{-6n}(T)-P_0}\right) 
	\nonumber\\
	&\quad =	F-\left(6n(n-1)t^{-2}(T)+T \right) F_T +12nt^{-1}(T)\,\left((1-n)\,t^{-1}(T)+\delta_1\sqrt{\frac{T}{6}}\right)T\,F_{TT}. \label{430}
\end{align}		
{The} possible solution to Equation \eqref{430} are by using a power-law ansatz $F(T) = F_{0}\,T^r$:
\begin{enumerate}
	\item {${\bf n=\frac{1}{2}}$:} We use the approximation $\frac{a_0^2\,\delta_1}{k}\sqrt{\frac{2T}{3}} \ll 1$, { the weak torsion scalar $T$ or far future $t(T) \gg 1$ approximations for analytic solutions, in agreement to Section \ref{sect23} characteristic equation solutions. This type of approximation will allow us to study the long-term cosmological universe evolution models in a simpler manner. By} setting the $+$ root, Equation \eqref{430} simplifies:
	\begin{align}
		&0 \approx C_1T^2 F+\left(2C_2C_1T^{7/2}-2C_2\sqrt{T}+C_2^2 T-1 \right) F_T
		+2C_2\left(C_2 T-\sqrt{T} \right)TF_{TT},  \label{431}
	\end{align}	
	where $C_1=\left(1+\frac{\rho_0^2}{P_0}\right) \frac{a_0^{6}}{216k^3}$ and $C_2=\frac{\delta_1 a_0^2}{2\sqrt{6}k}$. The solution of Equation \eqref{431} is
	\begin{align}
		& F(T) \approx F_{+}\,T^{r_{+}}+ F_{-}\,T^{r_{-}}   \label{431PL}
	\end{align}
	\begin{align}
		r_{\pm}=&\frac{1}{4C_2(C_2-1)}\Bigg[-\left(2C_1C_2-10C_2+7C_2^2-1\right) 
		\pm \Bigg[\left(2C_1C_2-10C_2+7C_2^2-1\right)^2
		\nonumber\\
		&\;-8C_2(C_2-1)\left(C_1(1-C_2)-\frac{25}{2}C_2+6C_2^2-3\right)\Bigg]^{1/2}\Bigg],
	\end{align}
	where $C_2 \neq \left\lbrace 0, 1 \right\rbrace$. { If $r_{\pm} \,\rightarrow\,1$, we will go to the TEGR limit.}
	
	\item {${\bf n=1}$:} For $C_3 \ll 1$, Equation \eqref{430} can be approximated as
	\begin{align}
		& 0\approx C_3\,T^2F-\frac{2}{C_4}\left(1+C_3\,T^3-\frac{C_4}{2}\right)F_T - \frac{2}{C_4}\,TF_{TT}, \label{432}
	\end{align}	
	where $C_3=\frac{\left(1+\frac{\rho_0^2}{P_0}\right)}{216\left(\frac{\delta\sqrt{-k}}{a_0}+1\right)^6}$ and $C_4=\frac{\delta\sqrt{-k}}{a_0}+1$. The solution of Equation \eqref{432} is described by Equation \eqref{431PL} with the $r_{\pm}$ roots:
	\begin{align}
		r_{\pm}=& -\frac{1}{2}\left(6+C_3-\frac{C_4}{2}\right) \pm \left[\frac{1}{4}\left(6+C_3-\frac{C_4}{2}\right)^2-\left(9-\frac{3C_4}{2}-\frac{C_3C_4}{2}\right)\right]^{1/2} .
	\end{align}
	{ If $r_{\pm} \,\rightarrow\,1$, we find the TEGR limit.}
	
	\item {${\bf n=2}$:} By using the approximation $\frac{\delta\,\delta_1\sqrt{-k}}{a_0\sqrt{6}}\sqrt{T} \ll 1$ { (far future approximation)} and setting the $+$ root, Equation \eqref{430} simplifies as
	\begin{align}
		& 0\approx C_5\,T^5\,F+\left(\frac{7}{2}+2C_5\,T^6\right)F_T +3TF_{TT}	,  	  \label{433}
	\end{align}	
	where $C_5\approx 5,233 \,\times \,10^{-9}\left(1+\frac{\rho_0^2}{P_0}\right) \ll 1$. The solution of Equation \eqref{432} is still Equation \eqref{431PL} with $r_{\pm}$ roots:
	\begin{align}\label{433a}
		r_{\pm}=& \frac{1}{12}\left[-\left(1+4C_5\right)\pm \left[1+16C_5^2-520C_5\right]^{1/2}\right] \approx \frac{1}{12}\left[-\left(1+4C_5\right)\pm \left(1-260C_5\right)\right].
	\end{align}
	Under the $C_5\,\rightarrow\, 0$ limit: $r_{+} \approx -22C_5\,\rightarrow\,0$ and $r_{-} \approx -\frac{1}{6}+\frac{64}{3}C_5\,\rightarrow\,-\frac{1}{6}$. In this case, Equation \eqref{431PL} leads to $F(T)\rightarrow -\Lambda_0+F_{-}\,T^{-1/6}$ with the cosmological constant $\Lambda_0$. This last solution is similar to those of ref. \cite{FTBcosmogholamilandry}. { The $r_{\pm} \,\rightarrow\,1$ TEGR limit is not possible under the weak $C_5$ approximation.}
	
	\item {${\bf n\,\rightarrow\,\infty}$:} Under this limit, Equation \eqref{430} simplifies as a simple homogeneous equation for $P_0 \neq -\rho_0^2$:
	\begin{align}
		&   T\,F_T  \approx \frac{F}{2} \quad\Rightarrow\quad F(T) \approx F_0\,\sqrt{T} , \label{434}
	\end{align}	
	{as we can find for any pure flat background cosmological simple solutions \cite{nonvacKSpaper,coleylandrygholami,scalarfieldTRW,myrzacosmotele1,myrzacosmospin,paliacosmo}.}
\end{enumerate}

\subsection{$k=+1$ Cosmological Solutions}\label{sect33}

Equation \eqref{1036} with $P(\rho)=\frac{P_0}{\rho}$ is solved by substituting the solution from Equation \eqref{400}:
\begin{align}
	&\left[-\frac{F}{2}+6n^2\,t^{-2}(T)\,F_T \right]\left(\frac{\left(P_0+\rho_0^2\right)\,t^{-6n}(T)+2P_0}{\left(P_0+\rho_0^2\right)\,t^{-6n}(T)-P_0}\right) 
	\nonumber\\
	&\quad =	F-\left( 6n(n-1)\,t^{-2}(T)+T
	\right)F_T+12n^2\,t^{-2}(T)\left(T-6n(n-1)t^{-2}(T)\right)\,F_{TT}. \label{460}
\end{align}
{The} possible solution to Equation \eqref{460} are by using the $F(T) = F_{0}\,T^r$ ansatz:
\begin{enumerate}
	\item {${\bf n=\frac{1}{2}}$:} By using the approximation $\frac{a_0^4}{6k^2}\,T\ll 1$ { (far future approximation)} and setting the $-$ root, Equation \eqref{460} simplifies as
	\begin{align}
		0\approx &C_7\,T^2 F+\left(3C_6\,T-2C_6C_7\,T^4-1\right)F_T+2C_6\left(1+C_6\,T\right)\,T^2F_{TT}, \label{461}
	\end{align}
	where $C_6=\frac{a_0^4}{24k^2}$ and $C_7=\left(1+\frac{\rho_0^2}{P_0}\right)\frac{a_0^6}{216k^3}\ll 1$. The solution of Equation \eqref{461} is Equation \eqref{431PL} with $r_{\pm}$ roots:
	\begin{align}
		r_{\pm} =& \frac{1}{4C_6(1+C_6)}\Bigg[-\left(13C_6-2C_6C_7+6C_6^2-1\right)\pm\Bigg[\left(13C_6-2C_6C_7+6C_6^2-1\right)^2
		\nonumber\\
		&\;-8C_6(1+C_6)\left(4C_6^2+21C_6-4\right)\Bigg]^{1/2}\Bigg],
	\end{align}
	where $C_6 \neq \left\lbrace -1,0\right\rbrace$. { For $r_{\pm} \,\rightarrow\,1$, we find the TEGR limit.}
	
	\item {${\bf n=1}$}: Equation \eqref{460} becomes
	\begin{align}
		0\approx &C_9\,T^2F+\left(1-\frac{2}{C_8} \left(1+C_9\,T^3\right)\right)F_T-\frac{2T\,F_{TT}}{C_8}, \label{462}
	\end{align}
	where $C_8=1-\frac{k}{a_0^2}$ and $C_9=\frac{\left(1+\frac{\rho_0^2}{P_0}\right)}{216\left(1-\frac{k}{a_0^2}\right)^3} \ll 1$. The solution of Equation \eqref{462} is Equation \eqref{431PL} with $r_{\pm}$ roots:
	\begin{align}
		r_{\pm} =& -\frac{1}{2}\left(C_9+\frac{C_8}{2}+6\right) \pm \left[\frac{1}{4}\left(C_9+\frac{C_8}{2}+6\right)^2-\left(9+\frac{3C_8}{2}-\frac{C_8C_9}{2}\right)\right]^{1/2} .
	\end{align}
	{ If $r_{\pm} \,\rightarrow\,1$, we find the TEGR limit.}
	
	\item {${\bf n=2}$:} By using the approximation $\frac{k}{24 a_0^2}\,T\ll 1$ and setting the $-$ root, Equation \eqref{460} simplifies under the $C_{10}\ll 1$ approximation as
	\begin{align}
		0\approx &C_{10}\,T^5F- \left(\frac{1}{2}+2C_{10}\,T^6\right)F_T-TF_{TT} , \label{463}
	\end{align}
	where $C_{10}=\left(1+\frac{\rho_0^2}{P_0}\right)\left(\frac{1}{24}\right)^6 \ll 1$. The solution of Equation \eqref{463} is Equation \eqref{431PL} with $r_{\pm}$ roots:
	\begin{align}
		r_{\pm} =& -\left(C_{10}+\frac{23}{4}\right) \pm \left[\left(C_{10}+\frac{23}{4}\right)^2-33+C_{10}\right]^{1/2} \approx -C_{10}-\frac{23}{4}\pm \left(\frac{1}{4}+25C_{10}\right).
	\end{align}
	Under the $C_{10}\,\rightarrow\,0$ limit: $r_{+}\,\rightarrow\,-\frac{11}{2}$ and $r_{-}\,\rightarrow\,-6$. Then Equation \eqref{431PL} becomes $F(T)\,\rightarrow\,F_{+}\,T^{-\frac{11}{2}}+F_{+}\,T^{-6}$. { The $r_{\pm} \,\rightarrow\,1$ TEGR limit is not possible under $C_{10}\ll 1$.}
	
	\item {${\bf n\,\rightarrow\,\infty}$:} Once again for $P_0 \neq -\rho_0^2$, we find under this limit the same differential equation and solution as Equation \eqref{434}, i.e., $F(T)\approx F_0\,\sqrt{T}$.
	
\end{enumerate}


\section{General Polytropic Gas Teleparallel Field Equation Solutions}\label{sect4}

\subsection{Conservation Law Solutions and Energy Conditions}
A polytropic gas is defined as $\rho=\rho(t)$ and $P=-P_0\,\rho^{1+\frac{1}{p}}$, where $\frac{1}{3}<P_0\leq 1$ and $0<p<\infty$ \cite{polytropic1,polytropic2,polytropic3,polytropic4,polytropic5,polytropic6,polytropic7,polytropic8}. Under the $p\,\rightarrow\,\infty$ limit, we find that $P=-P_0\,\rho$, a linear DE PF. However the $p\,\rightarrow\,0$ limit will lead to an infinitely huge pressure universe looking like the very early universe. Equation \eqref{1003} for this general type of gas is solved by using the $a(t)=a_0\,t^n$ ansatz:
\begin{align}\label{500}
	&\dot{\rho}+\frac{3n}{t}\,\rho\,\left(1-P_0\,\rho^{\frac{1}{p}}\right)=0,
	\nonumber\\
	&\Rightarrow\,\rho(t)=\left[P_0-\left(P_0-\rho_0^{-1/p}\right)t^{3n/p}\right]^{-p}.
\end{align}
{We} will apply the ECs defined by Equations \eqref{1040} for Equation \eqref{500} CL solution as
\begin{align}
	& \left(P_0-\rho_0^{-1/p}\right)t^{3n/p} \leq 0 , \quad \left(P_0-\rho_0^{-1/p}\right)t^{3n/p}\leq -2P_0 , \quad \left(P_0-\rho_0^{-1/p}\right)t^{3n/p} \leq P_0-|P_0|,  
	\nonumber\\
	& \left(P_0-\rho_0^{-1/p}\right)t^{3n/p} \leq P_0.   \label{500a}
\end{align}
{The} dominating energy conditions is exactly $t^{3n/p}\leq \frac{-2|P_0|}{\left(P_0-\rho_0^{-1/p}\right)}$. { Note that the dust matter limit (i.e., $P(\rho)\,\rightarrow\,0$ for any $\rho$) will be possible under the $P_0\,\rightarrow\,0$ limit as for the Chaplygin gas, but the $p\,\rightarrow\,\infty$ limit will allow us to obtain this limit by a closely linear PF. This case shows that the linear PF, in particular the $\Lambda$CDM model case, constitutes a sort of limit case when $p\,\rightarrow\,\infty$. The ECs defined by Equations \eqref{1040} are still more trivially satisfied under the ordinary matter limit.}

\subsection{$k=0$ Cosmological Solutions}\label{sect41}

The unified FEs using Equation \eqref{1012}, $P=-P_0\,\rho^{1+\frac{1}{p}}$ and Equation \eqref{500} are \cite{coleylandrygholami}
\begin{align} 
	& \frac{3}{2}\left(n-1+\frac{n\,P_0}{\left[{\beta}_p T^{-3n/2p}-P_0\right]}\right) G_T =  T\,G_{TT} ,
	\nonumber\\
	& \Rightarrow\;G_T(T)=G_T(0)\,T^{\frac{3}{2}(n-1)}\,\left[P_0\,T^{3n/2p}-\beta_p\right]^{-2p/3} , \label{501}
\end{align}
where ${\beta}_p=\left(P_0-\rho_0^{-1/p}\right)(6n^2)^{3n/2p}$. The Chaplygin gas case corresponds to the $p=-\frac{1}{2}$, i.e., $\beta_{1/2}=\beta$ in section \ref{sect3}. The general teleparallel $F(T)$ solution is exactly:
\begin{align}
	F(T)=& G(0)\sqrt{T}+G_T(0)\,\sqrt{T}\,\left[\int\,dT'\,T'^{3(n-1)/2}\,\left(P_0\,T^{3n/2p}-\beta_p\right)^{-2p/3}\right],
	\nonumber\\
	=& G(0)\sqrt{T}+G_T(0)\,\sqrt{T}\,I_p(T), \label{502}
\end{align}	
{where $I_p(T)$ is a new integral-based special function. Some values of $I_p(T)$ are displayed in Table \ref{table1}. The $p\geq 0$ solutions yield to hypergeometric functions, and $p<0$ solution yield to power-law superposition solutions. We confirm in part the refs \cite{coleylandrygholami,scalarfieldTRW} where $k=0$ flat cosmological case still yields to analytical teleparallel $F(T)$ solution for each value of $p$ and $n$.}

\begin{table}[ht]
\caption{Values of $I_p(T)$ function for Equation \eqref{502} teleparallel flat polytropic solutions.}
	\label{table1}
	\begin{tabular}{c|c}
		\hline
		$p$	& $I_p(T)$ \\
		\hline
		$0$ limit	& $\frac{2}{(3n-1)}\,T^{(3n-1)/2}$ \\
		\hline
		$\frac{3}{2}$	& $-\frac{2\,T^{(3n-1)/2}}{\beta_{3/2}\,(3n-1)}\,_2F_1\left(1,\frac{3n-1}{2n};\frac{5n-1}{2n};\frac{P_0}{\beta}\,T^n\right)$ \\
		\hline
		$3$	& $\frac{2\,T^{(3n-1)/2}}{\beta_{3}^2(3n-1)}\,_2F_1\left(2,\frac{3n-1}{n};\frac{4n-1}{n};\frac{P_0}{\beta_{3}}\,T^{n/2}\right)$ \\
		\hline
		$\frac{9}{2}$	& $-\frac{2\,T^{(3n-1)/2}}{\beta_{9/2}^3\,(3n-1)}\,_2F_1\left(3,\frac{3(3n-1)}{2n};\frac{11n-3}{2n};\frac{P_0}{\beta_{9/2}}\,T^{n/3}\right)$ \\
		\hline
	\end{tabular}
	
\end{table}

{ In the $n\,\rightarrow\,\infty$ limit (very fast universe expansion), we find from Equation \eqref{502} that $F(T)\approx G(0)\,\sqrt{T}+G_T(0)\,T^{3n/2}$ for any $p >0$ value as in ref. \cite{paliacosmo} for the dark torsion solution. For the TEGR limit, by setting $n=\frac{2}{3}$, $p\,\rightarrow\,0$ limit and $G(0) \ll G_T(0)$, we will find that Equation \eqref{502} goes to $F(T)\,\rightarrow\,T$ TEGR limit.}

\subsection{$k=-1$ Cosmological Solutions}\label{sect42}

Equation \eqref{1026} for $P=-P_0\,\rho^{1+\frac{1}{p}}$ is solved by substituting Equation \eqref{500} solution:
\begin{align}
	& \left[ -\frac{F}{2}+\sqrt{6}\,n\delta_1\,t^{-1}(T) \sqrt{T} F_T\right]\left(\frac{\left(P_0-\rho_0^{-1/p}\right)t^{3n/p}(T)+2P_0}{\left(P_0-\rho_0^{-1/p}\right)t^{3n/p}(T)-P_0}\right) 
	\nonumber\\
	&\quad =	F-\left(6n(n-1)t^{-2}(T)+T \right) F_T +12nt^{-1}(T)\,\left((1-n)\,t^{-1}(T)+\delta_1\sqrt{\frac{T}{6}}\right)T\,F_{TT}. \label{530}
\end{align}	
{The} possible solution to Equation \eqref{530} are by using the $F(T) = F_{0}\,T^r$ ansatz:
\begin{enumerate}
	\item {${\bf n=\frac{1}{2}}$:} {We will again use the weak torsion scalar approximation $\frac{a_0^2\,\delta_1}{k}\sqrt{\frac{2T}{3}} \ll 1$ (or far future $t(T) \gg 1$ case as shown in section \ref{sect23} characteristic equation solutions) as in sections \ref{sect32} and \ref{sect33} solutions. By setting} the $+$ root, Equation \eqref{530} becomes:
	\begin{align}
		0\approx	&  \frac{3F}{2}+\left(C_2^2 T+C_2 T^{1/2}-1 \right) TF_T
		+2C_2\left(C_2 \sqrt{T}-1 \right)T^{5/2}\,F_{TT},   \label{531}
	\end{align}	
	where $C_{11}=\left(1-\frac{1}{P_0\,\rho_0^{1/p}}\right)\,\frac{(-6k)^{3/2p}}{(\delta\delta_1 a_0)^{3/p}} \gg 1$ and then $\frac{C_{11}\,T^{-3/p}-2}{C_{11}\,T^{-3/p}+1} \,\rightarrow\, 1$ under the same limit. Using the same power-law ansatz, we find Equation \eqref{431PL} as solution with the roots:
	\begin{align}
		r_{\pm} = -\frac{1}{4C_2} \pm \left[\frac{1}{16C_2^2}-\frac{1+2C_2}{4C_2(C_2-1)}\right]^{1/2},
	\end{align}
	where $C_2 \neq \left\lbrace 0, 1 \right\rbrace$. { For $r_{\pm} \,\rightarrow\,1$: we find the TEGR limit.}
	
	\item {${\bf n=1}$:} We find
	\begin{align}
		0\approx	& \frac{3F}{2}-\left(1+\frac{1}{C_4}\right)T\,F_T + \frac{2}{C_4}\,T^{2}\,F_{TT}, \label{532}
	\end{align}	
	where $C_{12}=\left(1-\frac{1}{P_0\,\rho_0^{1/p}}\right)\,\left({\sqrt{6}\delta_1\left(\frac{\delta\sqrt{-k}}{a_0}+1\right)}\right)^{3/p}\gg 1$ and then $\frac{2+C_{12}T^{-3/2p}}{1-C_{12}T^{-3/2p}}\,\rightarrow\, -1$. We find Equation \eqref{431PL} as solution with the roots $r_{+}=\frac{C_4}{2}$ and $r_{-}=\frac{3}{2}$. { The TEGR limit is only possible for $C_4=2$ and $F_{+}\gg F_{-}$ in Equation \eqref{431PL}.}
	
	\item {${\bf n=2}$}: By using $\frac{\delta\,\delta_1\sqrt{-k}}{a_0\sqrt{6}}\sqrt{T} \ll 1$ { (far future approximation)} and the $+$ root, Equation \eqref{530} simplifies as
	\begin{align}
		0\approx	&  \frac{3F}{2}-\frac{1}{2} TF_T-3T^2\,F_{TT}, \label{533}
	\end{align}	
	where $C_{13}=\left(1-\frac{1}{P_0\,\rho_0^{1/p}}\right)\,(24)^{3/p} \gg 1$ and then $\frac{2+C_{13}T^{-3/p}}{1-C_{13}T^{-3/p}}\,\rightarrow\, -1$.  We find Equation \eqref{431PL} as solution with the roots $r_{\pm} = \frac{5 \pm \sqrt{97}}{12}$. { The TEGR limit is not possible.}

	\item ${\bf n\,\rightarrow\,\infty}$ limit: For $P_0 \neq \rho_0^{-1/p}$, we obtain Equation \eqref{434} with $F(T) \approx F_0\,\sqrt{T}$ as solution.
	
\end{enumerate}

\subsection{$k=+1$ Cosmological Solutions}\label{sect43}

Equation \eqref{1036} for $P=-P_0\,\rho^{1+\frac{1}{p}}$ is solved by substituting Equation \eqref{500} solution:
\begin{align}
	&\left[-\frac{F}{2}+6n^2\,t^{-2}(T)\,F_T \right]\left(\frac{\left(P_0-\rho_0^{-1/p}\right)t^{3n/p}(T)+2P_0}{\left(P_0-\rho_0^{-1/p}\right)t^{3n/p}(T)-P_0}\right) 
	\nonumber\\
	&\quad =	F-\left( 6n(n-1)\,t^{-2}(T)+T
	\right)F_T+12n^2\,t^{-2}(T)\left(T-6n(n-1)t^{-2}(T)\right)\,F_{TT}. \label{560}
\end{align}
The possible solution to Equation \eqref{560} are by using the $F(T) = F_{0}\,T^r$ ansatz:
\begin{enumerate}
	\item {${\bf n=\frac{1}{2}}$:} By using the approximation $\frac{a_0^4}{6k^2}\,T\ll 1$ { (far future approximation)} and the $-$ root, Equation \eqref{560} simplifies as
	\begin{align}
		0\approx	&  \frac{3F}{2} -TF_T+2C_6\left(1+C_6\,T\right)\,T^3F_{TT}, \label{561}
	\end{align}
	where $C_{14}=\left(1-\frac{1}{P_0\,\rho_0^{1/p}}\right)\,\left(-\frac{6k}{a_0^2}\right)^{3/2p} \gg 1$. We find Equation \eqref{431PL} as a solution with the roots:
	\begin{align}
		r_{\pm} = \frac{\left(1-2C_6+2C_6^2\right)}{4C_6(1+C_6)} \pm \left[\frac{\left(1-2C_6+2C_6^2\right)^2}{16C_6^2(1+C_6)^2}+\frac{1}{2}\right]^{1/2},
	\end{align}
	where $C_6 \neq \left\lbrace -1, 0 \right\rbrace$. { For $r_{\pm} \,\rightarrow\,1$: we find the TEGR limit.}
	
	\item {${\bf n=1}$:} With the approximation $\frac{a_0^4}{6k^2}\,T\ll 1$, we find that
	\begin{align}
		0\approx	&\frac{3F}{2}-\left(1+\frac{1}{C_8}\right)T\,F_T+\frac{2T^2\,F_{TT}}{C_8},\label{562}
	\end{align}
	where $C_{15}=\left(1-\frac{1}{P_0\,\rho_0^{1/p}}\right)\,\left(6\left(1-\frac{k}{a_0^2}\right)\right)^{3/2p} \gg 1$. We find Equation \eqref{431PL} as solution with the roots $r_{+}=\frac{C_8}{2}$ and $r_{-}=\frac{3}{2}$. { The TEGR limit is only possible for $C_8=2$ and $F_{+}\gg F_{-}$ in Equation \eqref{431PL}.}
	
	\item {${\bf n=2}$:} By using $\frac{k}{24 a_0^2}\,T\ll 1$, $C_{13} \gg 1$ { (far future approximation)} and $-$ root, Equation \eqref{560} simplifies as
	\begin{align}
		0\approx	&\frac{3F}{2}-\frac{5}{2}TF_T+T^2F_{TT} . \label{563}
	\end{align}
	We find Equation \eqref{431PL} as solution with the roots $r_{+}=3$ and $r_{-}=\frac{1}{2}$.
	
	\item ${\bf n\,\rightarrow\,\infty}$ limit: Here again for $P_0 \neq \rho_0^{-1/p}$, we obtain Equation \eqref{434} with $F(T) \approx F_0\,\sqrt{T}$ as solution.
\end{enumerate}

\section{Physical, Graphical and Experimental Comparisons of New Teleparallel \boldmath{$F(T)$} Gravity Solutions}\label{sect5}

\subsection{Comparison Between New Chaplygin and Polytropic Solutions and Consequences}

\begin{figure}[ht]
	\includegraphics[width=6.2cm,height=6.2cm]{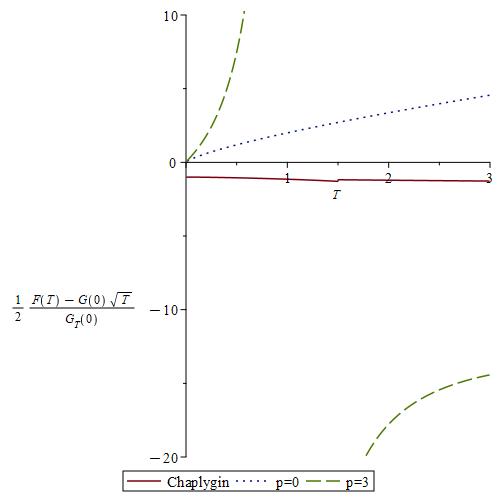} 
	\includegraphics[width=6.2cm,height=6.2cm]{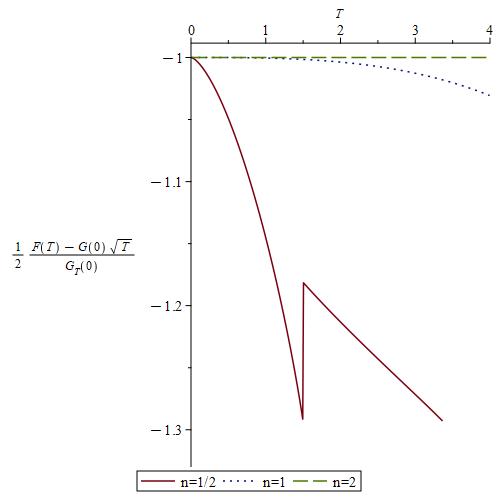}   \\
	\includegraphics[width=6.2cm,height=6.2cm]{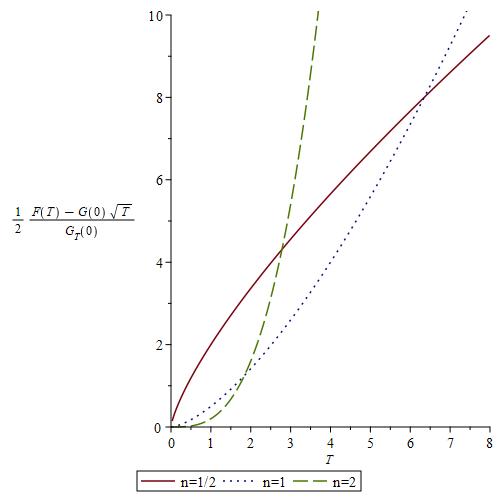}  
	\includegraphics[width=6.2cm,height=6.2cm]{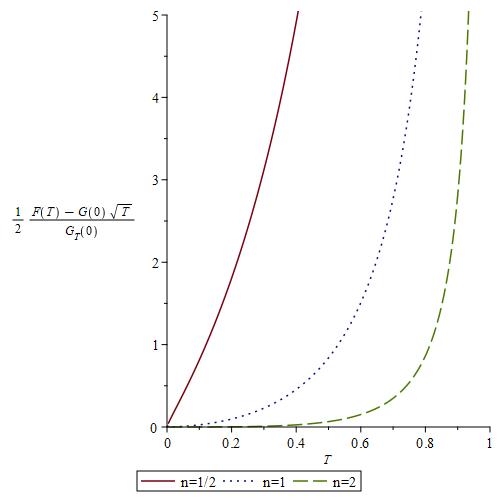} 
	\caption{Plot of flat cosmological $k=0$ teleparallel $F(T)$ solutions for Chaplygin and polytropic cosmological gas sources (top left: $n=\frac{1}{2}$ and any $p$, top right: Chaplygin, bottom left: $p\,\rightarrow\,0$ limit, bottom right: $p=3$).}
	\label{figure1}
\end{figure}

We compare the teleparallel $F(T)$ solutions for flat cosmological ($k=0$) Chaplygin and polytropic gases cases. We have plotted Equation \eqref{404} (Chaplygin gas), $p\,\rightarrow\,0$ and $3$ subcases of Equation \eqref{502} on  Figure \ref{figure1}. The top left subfigure compares for $n=\frac{1}{2}$ case the Chaplygin and polytropic $p$-type gases solutions. We removed the cosmological geometry (homogeneous) $G(0)\,\sqrt{T}$ term (i.e., the $n\,\rightarrow\,\infty$ solution) and kept only the source contribution for making the distinction between polytropic and Chaplygin gases for $n=\frac{1}{2}$, $1$ and $2$, { the slow, linear and fast universe expansion cases}. The Chaplygin gas { can be seen as} a specific case of the general polytropic gas, where we can consider the parameter $p=-\frac{1}{2}$ in the polytropic EoS (even if the EoS is defined for positive values of $p$). The Chaplygin gas case leads to some characteristic curves, which we can also see on the top-right graphs of figure \ref{figure1}. This feature makes the comparison more relevant and expresses the real nature of Chaplygin and polytropic gases. { We can imagine for larger $p$ cases the linear PF limit solutions leading to the $\Lambda$CDM ordinary matter limit for flat cosmological spacetimes. For larger $p$, the teleparallel $F(T)$ solutions will increase faster than for the $p\,\rightarrow\,0$ non-linear limit. The double power-law far future approximated teleparallel solutions become more suitable for cosmological evolution model studies, confirming the results of refs. \cite{myrzacosmotele1,myrzacosmospin,paliacosmo}. However, we will need in future works to determine the exact and finite non-linearity degree (or magnitude) by determining the parameter $p$ of a typical polytropic cosmological fluid model. Therefore, we can expect a high value of $p$-parameter close to infinity (linear PF limit), confirming a slightly non-linear $\Lambda$CDM model.}

For the spatially curved $k=\pm 1$ cases, we found some { far-future} approximated teleparallel $F(T)$ solutions, all described by double power-law defined by the form of Equation \eqref{431PL}. We only found the power $r_{\pm}$ in each cases after approximating Equations \eqref{430} and \eqref{460} for Chaplygin gas, and Equations \eqref{530} and \eqref{560} for general polytropic gas. A first feature is that the polytropic $p$-type teleparallel $F(T)$ solutions do not contain any $p$-dependent term, as seen in Sections \ref{sect42} and \ref{sect43} for $k=\pm 1$ solutions. The approximated polytropic teleparallel solutions are greatly simplified by parameter approximations { agreeing to far future cosmological evolution models, as specifically in ref. \cite{paliacosmo}. The TEGR limit ($r_{\pm}\,\rightarrow\,1$) is satisfied for most of subcases and for high-valued $p$-parameter values close to linear PF limit (and the $\Lambda$CDM limit too). In this sense, the $p=3$ subcase represents more closely the expected and realistic cosmological models, and solutions are more comparable to those of refs. \cite{coleylandrygholami,myrzacosmotele1,myrzacosmospin,paliacosmo}. The future research will have to explicitly determine the $p$-parameter for any of flat $k=0$ and non-flat $k=\pm 1$ cosmological models.}

A last consideration: In general, there is no cosmological constant $\Lambda_0$ in the { new} teleparallel solutions in almost all cases. The only exception to this rule are Equations \eqref{433} and \eqref{433a} under the $C_5\,\rightarrow\, 0$ limit of $n=2$ subcase, where $r_{+}=0$ leading to $F_{+}=-\Lambda_0$, and then a simple polynomial teleparallel $F(T)$ solution.

\subsection{Comparisons with Recent Experimental Results}

However, the new Chaplygin and polytropic teleparallel $F(T)$ solutions can be compared and tested with several experimental data sets from recent DE experiments such as Dark Energy Spectroscopic Instrument (DESI), and other Baryonic Data (BAO) and cosmological redshift measurements ($H(z)$-based measurements) \cite{DESI1,DESI2,DESI3,DESI4,DESI5,BAOvsSN1,BAO2}. The far future approximated teleparallel solution for $k=0$ cases can be easily compared to the mentioned DE data sets for baryonic as cosmological background types of data. The double power-law $k=\pm 1$ teleparallel $F(T)$ solutions for far future universe evolution should also be compared with data for more precise and most relevant non-flat cosmological models. Investigating from this point, we can determine which solutions are the most realistic for universe models and explanations.

Therefore, we can never forget that the current paper primarily aims to develop and find by theoretical and mathematical means the most relevant teleparallel $F(T)$ solutions for non-linear cosmological fluids such as Chaplygin and polytropic gases. In Sections \ref{sect3} and \ref{sect4} we find the most relevant and easily verifiable new teleparallel $F(T)$ solutions to be formally tested and compared with DESI or equivalent data sets. From this point, we can propose the main aim of future research to be comparing, by direct data fitting, these new teleparallel solutions with DE and background measurement as those made by the DESI collaboration \cite{DESI1,DESI2,DESI3,DESI4,DESI5}. We have seen that several of the new teleparallel $F(T)$ solutions are also close to those found in refs \cite{myrzacosmotele1,myrzacosmospin,paliacosmo}, especially Equation \eqref{431PL}-based double power-law $k=\pm 1$ solutions found in Sections \ref{sect32}--\ref{sect43}. These results will allow us to make good comparison with data sets in the future data fitting based works for determining which of the new solution classes are the most realistic for universe description.

Beyond the new teleparallel solution class determination, we will also be able to determine, by the same means, whether Chaplygin or polytropic fluid is more suitable for describing a DE-dominating universe with some amount of dust matter. We will be able to find the exact value of $p$ and $n$ parameters and also several other thermodynamic-based physical quantities for a DE dominating and expanding universe for a flat ($k=0$) and a non-flat ($k=\pm 1$) universe. We will be able to better confirm and/or adapt the $\Lambda$CDM models to the data sets and by determining the most suitable new teleparallel $F(T)$ solution models. Some recent similar studies using data comparison have been performed for more simple universe models cases using redshift $H(z)$, BAO and other similar data sets (see refs. \cite{celiaBAO1,celiaBAO2,celiaBAO3} and refs. within). We will be able to use the scale factor $a$ and determine the $n$-parameter possible values for the comparison with redshift measurements. Some data analysis techniques used in these mentioned works can be reused for future comparative works for the new teleparallel solutions found here. But the data comparison goes beyond the aims and scopes of the theoretical and mathematical physics based approach of the current paper. Nevertheless, the data fitting-based studies deserve to be performed in a near future and as suggested in some recent works for teleparallel $F(T,B)$-type extension solutions \cite{FTBcosmogholamilandry}. In the current work, we have all needed ingredients to achieve any future full data-based polytropic/Chaplygin fluid determination study.

\section{Concluding Remarks}\label{sectconcl}

We will first conclude the current paper by stating that the flat cosmological $k=0$ teleparallel $F(T)$ solutions are described by special functions, especially by hypergeometric functions plotted on the figure \ref{figure1} for Chaplygin ($p=-\frac{1}{2}$) and polytropic $p\,\rightarrow\,0$ limit and three gases. For the spatially curved cosmological $k=\pm 1$ teleparallel $F(T)$ solutions are practically all approximated by a double power-law function described by Equation \eqref{431PL}. The $k=\pm 1$ general polytropic solutions found in Sections \ref{sect42} and \ref{sect43} are not $p$-dependent because of some dominating terms approximations. There is an important feature in the new solutions: no cosmological constant $\Lambda_0$ term in almost all the cases. The only exception are Equations \eqref{433} and \eqref{433a} under the $C_5\,\rightarrow\,0$ limit, leading to a simple polynomial with $\Lambda_0$ term. Under all previous considerations, we can claim that non-flat $k=\pm 1$ teleparallel $F(T)$ solutions have some common points with those found for cosmological teleparallel $F(T)$ and $F(T,B)$ solutions found in refs \cite{coleylandrygholami,FTBcosmogholamilandry,scalarfieldTRW}, because we use the same coframe/spin-connection pair and ansatz. { However, the current paper is a starting point to verify and test the new teleparallel solutions on recent experimental data sets as performed in very recent studies with this aim \cite{celiaBAO1,celiaBAO2,celiaBAO3}. In the future, by using similar experimental data analysis approaches, we will need to formally compare the new solutions with DE, BAO and redshift experimental data sets. This will allow us to determine the most suitable classes of Chaplygin/polytropic teleparallel $F(T)$ solutions for explaining and determining the acceleration of expansion magnitude, the polytropic level and several other related physical parameters of the universe as precise as possible. 

}

Therefore, there are some perspectives for future research going beyond the current works. We can proceed, for example, with the same type of polytropic fluid sources for KS and static SS teleparallel spacetimes for new additional classes of solutions by using the same process and coframe ansatz approaches as in refs. \cite{staticscalarfieldSS,scalarfieldKS,nonvacSSpaper,nonvacKSpaper}. We can also use the same TRW geometry, replace polytropic sources by an electromagnetic source, proceed by the same approach, and finally find the corresponding classes of teleparallel $F(T)$ solutions. A last insight is that we will be able to study the Anti-deSitter (AdS) spacetimes in teleparallel gravity by using polytropic sources. However, we also need to investigate the polytropic principles in teleparallel gravity framework for possible symmetries and additional KVs. More physically, we will be able to elaborate a pure teleparallel quartessence models explaining the dark energy and the dark matter under an unified model. All these suggestions are feasible in the near future and will allow us to study the AdS wormholes and BHs solutions in teleparallel gravity.


%

\vspace{6pt} 


\begin{thebibliography}{999}
	
	\bibitem{Lucas_Obukhov_Pereira2009} Lucas, T.G.; Obukhov, Y.; Pereira, J.G. Regularizing role of teleparallelism. \textit{Phys. Rev. D} \textbf{2009}, \textit{80}, 064043. 
	
	\bibitem{Krssak:2018ywd} {Krššák}, M.;  {van~den~Hoogen}, {R.J.}; {Pereira}, {J.G.}; {Boehmer}, {C.G.}; {Coley}, {A.A.} Teleparallel Theories of Gravity: Illuminating a Fully Invariant Approach. \textit{Class. Quantum Gravity} \textbf{2019}, \textit{{36}}, {183001} .
	
	\bibitem{Bahamonde:2021gfp} Bahamonde, S.;  Dialektopoulos, K.F.;  Escamilla-Rivera, C.;  Farrugia, G.;  Gakis, V.;  Hendry, M.;  Hohmann, M.;  Said, J.L.;  Mifsud, J.;  Di Valentino, E. Teleparallel Gravity: From Theory to Cosmology. \textit{Rep. Prog. Phys.} \textbf{2023}, \textit{86}, 026901. 
	
	\bibitem{Krssak_Pereira2015} Krssak, M.; Pereira, J.G. Spin Connection and Renormalization of Teleparallel Action. \textit{Eur. Phys. J. C} \textbf{2015}, \textit{75}, 519.
	
	\bibitem{Coley:2019zld} Coley, {A.A.}; van den Hoogen, R.J.; McNutt, D.D. Symmetry and Equivalence in Teleparallel Gravity. \textit{J. Math. Phys.} \textbf{2020}, \textit{61}, 072503. 
	
	\bibitem{MCH} McNutt, D.D.;  Coley, A.A.;  van den Hoogen, R.J. A frame based approach to computing symmetries with non-trivial isotropy groups. \textit{J. Math. Phys.} \textbf{2023}, \textit{64}, 032503. 
	
	\bibitem{Aldrovandi_Pereira2013} Aldrovandi, R.;  Pereira, J.G. \textit{Teleparallel Gravity, An Introduction};  Springer: {Berlin/Heidelberg, Germany,} 2013.	
	
	\bibitem{olver1995equivalence} Olver, P. \textit{Equivalence, Invariants and Symmetry}; Cambridge University Press: Cambridge, UK, 1995.
	
	
	\bibitem{Krssak_Saridakis2015} {Krššák}, M.;  {Saridakis}, {E.N.} The covariant formulation of $f(T)$ gravity.  \textit{Class. Quantum Gravity} \textbf{2016}, \textit{33}, {115009}.
	
	\bibitem{Ferraro:2006jd} Ferraro, R.;  Fiorini, F. Modified teleparallel gravity: Inflation without an inflation. \textit{Phys. Rev. D} \textbf{2007}, \textit{75}, 084031.
	
	\bibitem{Ferraro:2008ey} Ferraro, R.;  Fiorini, F. On Born-Infeld Gravity in Weitzenbock spacetime. \textit{Phys. Rev. D} \textbf{2008}, \textit{78}, 124019. 
	
	\bibitem{Linder:2010py} Linder, E. Einstein's Other Gravity and the Acceleration of the Universe. \textit{Phys. Rev. D} \textbf{2010}, \textit{81}, 127301; Erratum in \textit{Phys. Rev. D} \textbf{2010}, \textit{82}, 109902.
	
	\bibitem{kayashi} Hayashi, K.;   Shirafuji, T. New general relativity. \textit{Phys. Rev. D} \textbf{1979}, \textit{19}, 3524.
	
	\bibitem{beltranngr} Jimenez, J.B.;   Dialektopoulos, K.F. Non-Linear Obstructions for Consistent New General Relativity. \textit{J. Cosmol. Atroparticle Phys.} \textbf{2020}, \emph{2020}, 18. 
	
	\bibitem{bahamondengr} Bahamonde, S.;  Blixt, D.;  Dialektopoulos, K.F.; Hell A.  Revisiting Stability in New General Relativity. \emph{Phys. Rev. D} \textbf{2024}, \emph{111}, 064080. 
	
	\bibitem{heisenberg1} Heisenberg, L. Review on $f(Q)$ Gravit. \emph{Phys. Rep.} \textbf{2023}{, \emph{1066},} 
1--78. 
	
	\bibitem{heisenberg2} Heisenberg, L.; Hohmann, M.;   Kuhn, S.  Cosmological teleparallel perturbations, \emph{J. Cosmol. Astropart. Phys.} \textbf{2024}, \emph{3}, 63. 
	
	\bibitem{faithman1} Flathmann, K.;   Hohmann, M. Parametrized post-Newtonian limit of generalized scalar-nonmetricity theories of gravity. \textit{Phys. Rev. D} \textbf{2022}, \textit{105}, 044002. 
	
	\bibitem{hohmannfq} Hohmann, M. General covariant symmetric teleparallel cosmology. \textit{Phys. Rev. D} \textbf{2021}, \textit{104}, 124077. 
	
	\bibitem{jimeneztrinity} Jimenez, J.B.; Heisenberg, L.; Koivisto, T.S. The Geometrical Trinity of Gravity. \textit{Universe} \textbf{2019}, \textit{5}, 173. 
	
	\bibitem{nakayama} Nakayama, Y. Geometrical trinity of unimodular gravity. \textit{Class. Quantum Gravity} \textbf{2023}, \textit{40}, 125005. 
	
	\bibitem{ftqgravity} Xu, Y.; Li, G.; Harko, T.; Liang, S.-D.  $f(Q,T)$ gravity. \textit{Eur. Phys. J. C} \textbf{2019}, \textit{79}, 708. 
	
	\bibitem{frtspecial} Maurya, D.C.;   Myrzakulov, R.;  Exact Cosmology in Myrzakulov Gravity, \textit{ Eur. Phys. J. C} \textbf{2024}, 84, 625. 
	
	\bibitem{frttheory} Harko, T.; Lobo, F.S.N.; Nojiri, S.; Odintsov, S.D. $f(R,T)$ gravity. \textit{Phys. Rev. D} \textbf{2011}, \textit{84}, 024020.
	
	\bibitem{myrzakulov1} Momeni, D.;  Myrzakulov, R. Myrzakulov Gravity in Vielbein Formalism: A Study in Weitzenböck Spacetime. \textit{Nucl. Phys. B} \textbf{2025}, \emph{1015}, 116903. 
	
	\bibitem{myrzakulov2} Maurya, D.C.; Yesmakhanova, K.; Myrzakulov, R.; Nugmanova, G. Myrzakulov $F(T,Q)$ gravity: Cosmological implications and constraints. \textit{Phys. Scr.} \textbf{2024}, \textit{99}, 10. 
	
	\bibitem{myrzakulov3} Maurya, D.C.; Yesmakhanova, K.; Myrzakulov, R.; Nugmanova, G. FLRW Cosmology in Metric-Affine $F(R,Q)$ Gravity. \textit{Chin. Phys. C} \textbf{2024}, \textit{48}, 125101. 
	
	\bibitem{myrzakulov4} Maurya D.C.; Myrzakulov, R. Transit cosmological models in Myrzakulov F(R,T) gravity theory. \textit{Eur. Phys. J. C} \textbf{2024}, \textit{84}, 534. 
	
	\bibitem{myrzakulov5} Mandal, S.; Myrzakulov, N.; Sahoo, P.K.; Myrzakulov, R. Cosmological bouncing scenarios in symmetric teleparallel gravity. \textit{Eur. Phys. J. Plus} \textbf{2021}, \textit{136}, 760. 
	
	\bibitem{golov1} Golovnev, A.; Guzman, M.-J. Approaches to spherically symmetric solutions in $f(T)$-gravity. \textit{Universe} \textbf{2021}, \textit{7}, 121. 
	
	\bibitem{golov2} Golovnev, A. Issues of Lorentz-invariance in $f(T)$-gravity and calculations for spherically symmetric solutions. \textit{Class. Quantum Gravity} \textbf{2021}, \textit{38}, 197001. 
	
	\bibitem{debenedictis} DeBenedictis, A.; Iliji\'c, S.;  Sossich, M. On spherically symmetric vacuum solutions and horizons in covariant $f(T)$ gravity theory. \textit{Phys. Rev. D} \textbf{2022}, \textit{105}, 084020. 
	
	\bibitem{baha1} Bahamonde, S.;  Camci, U. Exact Spherically Symmetric Solutions in Modified Teleparallel gravity. \textit{Symmetry} \textbf{2019}, \textit{11}, 1462. 
	
	\bibitem{awad1} Awad, A.; Golovnev, A.; Guzman, M.-J.;  El Hanafy, W. Revisiting diagonal tetrads: New Black Hole solutions in $f(T)$-gravity. \textit{Eur. Phys. J. C} \textbf{2022}, \textit{82}, 972. 
	
	\bibitem{bahagolov1} Bahamonde, S.; Golovnev, A.; Guzm\'an, M.-J.; Said, J.L.;  Pfeifer, C. Black Holes in $f(T,B)$ Gravity: Exact and Perturbed Solutions. \textit{J. Cosmol. Atroparticle Phys.} \textbf{2022}, \textit{1}, 037. 
	
	\bibitem{baha6} Bahamonde, S.; Faraji, S.; Hackmann, E.;  Pfeifer, C. Thick accretion disk configurations in the Born-Infeld teleparallel gravity. \textit{Phys. Rev. D} \textbf{2022}, \textit{106}, 084046.
	
	\bibitem{nashed5} Nashed, G.G.L. Quadratic and cubic spherically symmetric black holes in the modified teleparallel equivalent of general relativity: Energy and thermodynamics. \textit{Class. Quantum Gravity} \textbf{2021}, \textit{38}, 125004. 
	
	\bibitem{pfeifer2} Pfeifer, C.; Schuster, S. Static spherically symmetric black holes in weak $f(T)$-gravity. \textit{Universe} \textbf{2021}, \textit{7}, 153. 
	
	\bibitem{elhanafy1} El Hanafy, W.; Nashed, G.G.L. Exact Teleparallel Gravity of Binary Black Holes. \textit{Astrophys. Space Sci.} \textbf{2016}, \textit{361}, 68. 
	
	\bibitem{benedictis3} Aftergood, J.;  DeBenedictis, A. Matter Conditions for Regular Black Holes in $f(T)$ Gravity. \textit{Phys. Rev. D} \textbf{2014}, \textit{90}, 124006. 
	
	\bibitem{baha10} Bahamonde, S.; Doneva, D.D.; Ducobu, L.; Pfeifer, C.;  Yazadjiev, S.S. Spontaneous Scalarization of Black Holes in Gauss-Bonnet Teleparallel Gravity. \textit{Phys. Rev. D} \textbf{2023}, \textit{107}, 104013. 
	
	\bibitem{baha4} Bahamonde, S.; Ducobu, L.;  Pfeifer, C. Scalarized Black Holes in Teleparallel Gravity. \textit{J. Cosmol. Atroparticle Phys.} \textbf{2022}, {\textit{2022}}, 18. 
	
	\bibitem{ruggiero2} Iorio, L.; Radicella, N.;   Ruggiero, M.L. Constraining $f(T)$ gravity in the Solar System. \textit{J. Cosmol. Atroparticle Phys.} \textbf{2015}, \textit{{2015}}, 21. 
	
	\bibitem{sahoo1} Pradhan, S.; Bhar, P.; Mandal, S.; Sahoo, P.K.;  Bamba, K. The Stability of Anisotropic Compact Stars Influenced by Dark Matter under Teleparallel Gravity: An Extended Gravitational Deformation Approach. \textit{Eur. Phys. J. C} \textbf{2025}, \textit{85}, 127. 
	
	\bibitem{sahoo2} Mohanty, D.; Ghosh, S.;  Sahoo, P.K. Charged gravastar model in noncommutative geometry under $f(T)$ gravity. \textit{Phys. Dark Universe} \textbf{2025}, \textit{46}, 101692. 
	
	\bibitem{calza} Calza, M.;   Sebastiani, L. A class of static spherically symmetric solutions in $f(T)$-gravity. \textit{Eur. Phys. J. C} \textbf{2024}, \textit{84}, 476. 
	
	\bibitem{SSpaper} Coley, A.A.;  Landry, A.;  van den Hoogen, R.J.;  McNutt, D.D.  Spherically symmetric teleparallel geometries. \textit{Eur. Phys. J. C} \textbf{2024}, \textit{84}, 334. 
	
	\bibitem{nonvacSSpaper} Landry, A. Static spherically symmetric perfect fluid solutions in teleparallel $F(T)$ gravity. \textit{Axioms} \textbf{2024}, \textit{13}, 333.
	
	\bibitem{nonvacKSpaper} Landry, A. Kantowski-Sachs spherically symmetric solutions in teleparallel $F(T)$ gravity. \textit{Symmetry} \textbf{2024} \textit{16}, 953.
	
	\bibitem{roberthudsonSSpaper} van den Hoogen, R.J.;  Forance, H. Teleparallel Geometry with Spherical Symmetry: The diagonal and proper frames. \textit{J. Cosmol. Astrophys.} \textbf{2024}, \emph{11}, 033. 
	
	\bibitem{scalarfieldKS} Landry, A. Scalar field Kantowski-Sachs spacetime solutions in teleparallel $F(T)$ gravity. \textit{Universe} \textbf{2025}, \emph{11}, 26. 
	
	\bibitem{staticscalarfieldSS} Landry, A. Scalar Field Static Spherically Symmetric Solutions in Teleparallel $F(T)$ Gravity. \textit{Mathematics} \textbf{2025}, \emph{13}, 1003.
	
	\bibitem{TdSpaper} Coley, A.A.;  Landry, A.;  van den Hoogen, R.J.;  McNutt, D.D. Generalized Teleparallel de Sitter geometries. \textit{Eur. Phys. J. C} \textbf{2023}, \textit{83}, 977. 
	
	\bibitem{golov3} Golovnev, A.;  Guzman, M.-J.  Bianchi identities in $f(T)$-gravity: Paving the way to confrontation with astrophysics. \textit{Phys. Lett. B} \textbf{2020}, \textit{810}, 135806. 
	
	\bibitem{scalarfieldTRW} Landry, A.  Scalar field source Teleparallel Robertson-Walker $F(T)$-gravity solutions. \textit{Mathematics} \textbf{2025}, \textit{13}, 374. 
	
	\bibitem{coleylandrygholami} Coley, A.A.;  Landry, A.;  Gholami, F.  Teleparallel Robertson-Walker Geometries and Applications. \textit{Universe} \textbf{2023}, \textit{9}, 454. 
	
	\bibitem{preprint} Coley, A.A.;  van den Hoogen, R.J.;  McNutt, D.D.  Symmetric Teleparallel Geometries. \textit{Class. Quantum Gravity} \textbf{2022}, \textit{39}, 22LT01. 
	
	\bibitem{aldrovandi2003} Aldrovandi, R.;  Cuzinatto, R.R.;  Medeiros, L.G.  Analytic solutions for the $\Lambda$-FRW Model, \textit{Found. Phys.} \textbf{2006}, \textit{36}, 1736--1752. 
	
	\bibitem{bounce} Casalino, A.;  Sanna, B.;  Sebastiani, L.;  Zerbini, S.  Bounce Models within Teleparallel modified gravity,  \textit{Phys. Rev. D} \textbf{2021}, \textit{103}, {023514}.
	
	\bibitem{Capozz} Capozziello, S.;  Luongo, O.;  Pincak, R.;  Ravanpak, A.  Cosmic acceleration in non-flat $f(T)$ cosmology, \textit{Gen. Relativ. Gravit.} \textbf{2018}, \textit{50}, 53. 
	
	\bibitem{inflat} Bahamonde, S.;  Dialektopoulos, K.F.;  Hohmann, M.;  Said, J.L.;  Pfeifer, C.;  Saridakis, E.N. Perturbations in Non-Flat Cosmology for $f(T)$ gravity. \textit{Eur. Phys. J. C} \textbf{2023}, \textit{83}, 193. 
	
	\bibitem{FTBcosmogholamilandry} Gholami, F.;  Landry, A.  Cosmological solutions in teleparallel $F(T,B)$ gravity. \textit{Symmetry} \textbf{2025}, \textit{17}, 60. 
	
	\bibitem{HJKP2018} Hohmann, M.;  Järv, L.;  Krššák, M.; Pfeifer, C.  Modified teleparallel theories of gravity in symmetric spacetimes. \textit{Phys. Rev. D} \textbf{2019} \textit{100}, {084002}. 
	
	\bibitem{Cai_2015}	Cai, Y.-F.;  Capozziello, S.;  De Laurentis, M.;  Saridakis, E.N.  $f(T)$ teleparallel gravity and cosmology. \textit{Rep. Prog. Phys.} \textbf{2016}, \textit{79}, 106901. 
	
	\bibitem{dixit} Dixit, A.; Pradhan, A. Bulk Viscous Flat FLRW Model with Observational Constraints in $f(T,B)$ Gravity. \textit{Universe} \textbf{2022}, \textit{8}, 650.
	
	\bibitem{ftbcosmo3} Chokyi, K.K.;  Chattopadhyay, S.  Cosmological Models within $f(T,B)$ Gravity in a Holographic Framework. \textit{Particles} \textbf{2024}, \textit{7}, 856.
	
	
	\bibitem{hawkingellis1} Hawking, S.W.;  Ellis, G.F.R.  \textit{The Large Scale Structure of Space-Time}; Cambridge University Press: {Cambridge, UK,} 2010.
	
	\bibitem{cosmofluidsbohmer} Bohmer, C.G.;  d’Alfonso del Sordo, A.  Cosmological fluids with boundary term couplings, \textit{Gen. Relativ. Gravit.} \textbf{2024}, \textit{56}, 75. 
	
	\bibitem{BahamondeBohmer} Bahamonde, S.;  Bohmer, C.G.;  Carloni, S.;  Copeland, E.J.;  Fang, W.;  Tamanini, N.  Dynamical systems applied to cosmology: Dark energy and modified gravity. \textit{Phys. Rep.} \textbf{2018}, \textit{775-777}, 1--122. 
	
	\bibitem{steinhardt1} Zlatev, I.;  Wang, L.;  Steinhardt, P.  Quintessence, Cosmic Coincidence, and the Cosmological Constant. \textit{Phys. Rev. Lett.} \textbf{1999}, \textit{82}, 896. 
	
	\bibitem{steinhardt2} Steinhardt, P.;  Wang, L.;  Zlatev, I. Cosmological tracking solutions. \textit{Phys. Rev. D} \textbf{1999}, \textit{59}, 123504. 
	
	\bibitem{steinhardt3} Caldwell, R.R.;  Dave, R.;  Steinhardt, P. Cosmological Imprint of an Energy Component with General Equation of State. \textit{Phys. Rev. Lett.} \textbf{1998}, \textit{80}, 1582. 
	
	\bibitem{carroll1} Carroll, S.M. Quintessence and the Rest of the World. \textit{Phys. Rev. Lett.} \textbf{1998}, \textit{81}, 3067. 
	
	\bibitem{quintessencecmbpeak} Doran, M.;  Lilley, M.;  Schwindt, J.;  Wetterich, C. Quintessence and the Separation of CMB Peaks, \textit{Astrophys. J.} \textbf{2001}, \textit{559}, 501. 
	
	\bibitem{quintessenceholo} Zeng, X.-X.;  Chen, D.-Y.;  Li, L.-F. Holographic thermalization and gravitational collapse in the spacetime dominated by quintessence dark energy. \textit{Phys. Rev. D}, \textbf{2015}, \textit{91}, 046005. 
	
	\bibitem{quintchakra2024} Chakraborty, S.;  Mishra, S.;  Chakraborty, S. Dynamical system analysis of quintessence dark energy model. \textit{Int. J. Geom. Methods Mod. Phys.} \textbf{2025}, \textit{22}, 2450250.
	
	\bibitem{steinhardt2024} Shlivko, D.; Steinhardt, P.J.  Assessing observational constraints on dark energy, \textit{Phys. Lett. B} \textbf{2024}, \textit{855}, 138826. 
	
	\bibitem{rollingscalarfield} Ratra, B.; Peebles, P.J.E. Cosmological consequences of a rolling homogeneous scalar field. \textit{Phys. Rev. D} \textbf{1988}, \textit{37}, 3406.
	
	\bibitem{wolf1} Wolf, W.J.;  Ferreira, P.G. Underdetermination of dark energy. \textit{Phys. Rev. D} \textbf{2023}, \textit{108}, 103519. 
	
	\bibitem{wolf2} Wolf, W.J.; García-García, C.; Bartlett, D.J.;  Ferreira, P.G. Scant evidence for thawing quintessence. \textit{Phys. Rev. D} \textbf{2024}, \textit{110}, 083528. 
	
	\bibitem{wolf3}  Wolf, W.J.; Ferreira, P.G.;  García-García, C. Matching current observational constraints with nonminimally coupled dark energy. \textit{Phys. Rev. D} \textbf{2025},  \emph{111}, L041303.
	
	
	\bibitem{cosmofate} Wetterich, C. Cosmology and the Fate of Dilatation Symmetry. \textit{Nucl. Phys. B} \textbf{1988}, \textit{302}, 668. 
	
	\bibitem{quintessencephantom} Chiba, T.;  Okabe, T.;  Yamaguchi, M. Kinetically Driven Quintessence, \textit{Phys. Rev. D} \textbf{2000}, \textit{62}, 023511.
	
	\bibitem{strongnegative} Carroll, S.M.;  Hoffman, M.;  Trodden, M. Can the dark energy equation-of-state parameter w be less than -1? \textit{Phys. Rev. D} \textbf{2003}, \textit{68}, 023509.
	
	\bibitem{caldwell1} Caldwell, R.R. A phantom menace? Cosmological consequences of a dark energy component with super-negative equation of state. \textit{Phys. Lett. B} \textbf{2002}, \textit{545}, 23. 
	
	\bibitem{farnes} Farnes, J.S. A Unifying Theory of Dark Energy and Dark Matter: Negative Masses and Matter Creation within a Modified $\Lambda$CDM Framework. \textit{Astron. Astrophys.} \textbf{2018}, \textit{620}, A92. 
	
	\bibitem{baumframpton} Baum, L.;  Frampton, P.H. Turnaround in Cyclic Cosmology. \textit{Phys. Rev. Lett.} \textbf{2007}, \textit{98}, 071301. 
	
	\bibitem{phantomdivide} Hu, W. Crossing the Phantom Divide: Dark Energy Internal Degrees of Freedom. \textit{Phys. Rev. D} \textbf{2005}, \textit{71}, 047301. 
	
	\bibitem{phantomteleparallel1} Karimzadeh, S.;  Shojaee, R. Phantom-Like Behavior in Modified Teleparallel Gravity. \textit{Adv. High Energy Phys.} \textbf{2019}, {\emph{2019},} 4026856. 
	
	\bibitem{ripphantomteleparallel2} Pati, L.;  Kadam, S.A.;  Tripathy, S.K.;  Mishra, B. Rip cosmological models in extended symmetric teleparallel gravity. \textit{Phys. Dark Universe} \textbf{2022}, \textit{35}, 100925. 
	
	\bibitem{phantomteleparallel3} Kucukakca, Y.;  Akbarieh, A.R.;  Ashrafi, S.  Exact solutions in teleparallel dark energy model. \textit{Chin. J. Phys.} \textbf{2023}, \textit{82}, 47.
	
	\bibitem{quintom1} Cai, Y.-F.;  Saridakis, E.N.;  Setare, M.R.;  Xia, J.-Q. Quintom Cosmology: Theoretical implications and observations. \textit{Phys. Rep.} \textbf{2010}, \textit{493}, 1. 
	
	\bibitem{quintom2} Guo, Z.-K.;  Piao, Y.-S.;  Zhang, X.;  Zhang, Y.-Z. Cosmological evolution of a quintom model of dark energy. \textit{Phys. Lett. B} \textbf{2005}, \textit{608}, 177. 
	
	\bibitem{quintom3} Feng, B.;  Li, M.;  Piao, Y.-S.;  Zhang, X. Oscillating quintom and the recurrent universe. \textit{Phys. Lett. B} \textbf{2006}, \textit{634}, 101.
	
	\bibitem{quintom4} Mishra, S.;  Chakraborty, S. Dynamical system analysis of quintom dark energy model. \textit{Eur. Phys. J. C} \textbf{2018}, \textit{78}, 917. 
	
	\bibitem{quintomcoleytot} Tot, J.;  Coley, A.A.;  Yildrim, B.;  Leon, G. The dynamics of scalar-field Quintom cosmological models. \textit{Phys. Dark Universe} \textbf{2023}, \textit{39}, 101155.
	
	\bibitem{quintomteleparallel1} Bahamonde, S.;  Marciu, M.;  Rudra, P. Generalised teleparallel quintom dark energy non-minimally coupled with the scalar torsion and a boundary term. \textit{J. Cosmol. Astropart. Phys.} \textbf{2018}, \textit{4}, 56. 
	
		\bibitem{myrzacosmotele1} Myrzakulov, R. Accelerating universe from $F(T)$ gravity. \textit{Eur. Phys. J. C} \textbf{2011}, \textit{71}, 1752. 
		
		\bibitem{myrzacosmospin} Myrzakulov, R.;  Saez-Gomez, D.;  Tsyba, P.  Cosmological solutions in $F(T)$ gravity with the presence of spinor fields. \textit{Int. J. Geom. Methods Mod. Phys.} \textbf{2015}, \textit{12}, 1550023.
		
		\bibitem{paliacosmo} Paliathanasis, A.  $F(T)$ Cosmology with Nonzero Curvature. \emph{Mod. Phys. Lett. A} \textbf{2021}, \textit{36}, 2150261.
	
	\bibitem{Kamenshchik2001} Gorini, V.;  Kamenshchik, A.Y.;  Moschella, U.;  Pasquier, V. Chaplygin gas as a model for dark energy. In Proceedings of the {Tenth Marcel Grossmann Meeting: On Recent Developments in Theoretical and Experimental General Relativity, Gravitation and Relativistic Field Theories (In 3 Volumes), Rio de Janeiro, Brazil, 20--26 July} 2003; p. 840.
	
	\bibitem{Bento2002} Bento, M.C.;  Bertolami, O.; Sen, A.A. Generalized Chaplygin Gas Model: Dark Energy-Dark Matter Unification and CMBR Constraints. \textit{Gen. Relativ. Gravit.} \textbf{2003}, \textit{35}, 2063. 
	
	\bibitem{Bilic:2002chg} Bilić, N.;  Tuppe, G.B.;  Viollier, R.D. Unification of dark matter and dark energy: The inhomogeneous Chaplygin gas. \textit{Phys. Lett. B} \textbf{2002}, \emph{535}, 17. 
	
	\bibitem{Makler:2003iw}  Makler, M.;  de Oliveira, S.Q.;  Waga, I. Observational constraints on Chaplygin quartessence: Background results. \textit{Phys. Rev. D} \textbf{2003}, \emph{68}, 123521. 
	
	\bibitem{zhu2004} Zhu, Z.-H. Generalized Chaplygin gas as a unified scenario of dark matter/energy: Observational constraints. \textit{Astron. Astrophys.} \textbf{2004}, \emph{423}, 421. 
	
	\bibitem{polytropic1} Karami, K.;  Ghaffari, S.;  Fehri, J. Interacting polytropic gas model of phantom dark energy in non-flat universe. \textit{Eur. Phys. J. C} \textbf{2009}, \emph{64}, 85. 
	
	\bibitem{polytropic2} Karami, K.;  Safari, Z.;  Asadzadeh, S. Cosmological constraints on polytropic gas model. \textit{Int. J. Theor. Phys.} \textbf{2014}, \emph{53}, 1248. 
	
	\bibitem{polytropic3} Karami, K.;  Abdolmaleki, A. Reconstructing interacting new agegraphic polytropic gas model in non-flat FRW universe. \textit{Astrophys. Space Sci.} \textbf{2010}, 330, 133. 
	
	\bibitem{polytropic4} Karami, K.;  Khaledian, M.S. Polytropic and Chaplygin $f(R)$-gravity models. \textit{Int. J. Mod. Phys. D} \textbf{2012}, \emph{21}, 1250083. 
	
	\bibitem{polytropic5} Banerjee, S.;  Paul, A. Effect of Accretion on the evolution of Primordial Black Holes in the context of Modified Gravity Theories. {\emph{arXiv}} \textbf{2024}, arXiv:2406.04605. 
	
	\bibitem{polytropic6} Aboueisha, M.S.;  Nouh, M.I.;  Abdel-Salam, E.A-B.; Kamel, T.M.;  Beheary, M.M.;  Gadallah, K.A.K. Analysis of the Fractional Relativistic Polytropic Gas Sphere. \textit{Sci. Rep.} \textbf{2023}, \emph{13}, 14304.
	
	\bibitem{polytropic7} Cardenas, V.H.;  Cruz, M. Emulating dark energy models with known equation of state via the created cold dark matter scenario, \textit{Phys. Dark Universe} \textbf{2024}, \emph{44}, 101452. 
	
	\bibitem{polytropic8} Jia, Y.;  He, T.-Y.;  Wang, W.-Q.; Han, Z.-W.;  Yang, R.-J. Accretion of matter by a Charged dilaton black hole. { \emph{Eur. Phys. J. C}} \textbf{2024}, \emph{84}, 501. 
	
	\bibitem{Arun:2017dm} Arun, K.;  Gudennavar, S.B.;  Sivaram, C. Dark matter, dark energy, and alternate models: A review. \textit{Adv. Space Res.} \textbf{2017}, \textbf{60}, 166. 
	
	\bibitem{hypermomentum1} Iosifidis, D. Cosmological Hyperfluids, Torsion and Non-metricity. \textit{Eur. Phys. J. C} \textbf{2020}, \textit{80}, 1042.
	
	\bibitem{hypermomentum2} Heisenberg, L.;  Hohmann, M.;  Kuhn, S. Homogeneous and isotropic cosmology in general teleparallel gravity. \textit{Eur. Phys. J. C} \textbf{2023}, \textit{83}, 315. 
	
	\bibitem{hypermomentum3} Heisenberg, L.;  Hohmann, M. Gauge-invariant cosmological perturbations in general teleparallel gravity. \textit{Eur. Phys. J. C} \textbf{2024}, \textit{84}, 462.
	
	\bibitem{Kontou:2020bta} Kontou, E.-A.;  Sanders, K. Energy conditions in general relativity and quantum field theory. \textit{Class. Quantum Gravity} \textbf{2020} \textit{37}, 193001.
	

	\bibitem{DESI1} DESI Collaboration. The Dark Energy Survey: Cosmology Results With 1500 New High-redshift Type Ia Supernovae
	Using The Full 5-year Dataset. \emph{ Astrophys. J. Lett.} \textbf{2024}, \textit{973}, L14. 
	
	\bibitem{DESI2} DESI Collaboration. DESI 2024 VI: Cosmological Constraints from the Measurements of Baryon Acoustic Oscillations. \emph{J. Cosmol. Astropart. Phys.} \textbf{2025}, \emph{2}, 21. 
	
	\bibitem{DESI3} DESI Collaboration. Dark Energy Survey: Implications for cosmological expansion models from the final DES Baryon
	Acoustic Oscillation and Supernova data. \emph{arXiv} \textbf{2025}, arXiv:2503.06712. 
	
	\bibitem{DESI4} DESI Collaboration. DESI DR2 Results II: Measurements of Baryon Acoustic Oscillations and
	Cosmological Constraint. \emph{arXiv} \textbf{2025},  arXiv:2503.14738. 
	
	\bibitem{DESI5} DESI Collaboration. DESI DR2 Results I: Baryon Acoustic Oscillations from the Lyman Alpha Forest. \emph{Phys. Rev. D} \textbf{2025}, {\emph{in press}}. 
	
	\bibitem{BAOvsSN1} Notari, A.;  Redi, M.;  Tesi, A. BAO vs. SN evidence for evolving dark energy. \emph{J. Cosmol. Astropart. Phys.} \textbf{2025}, \emph{4}, 48. 
	
	\bibitem{BAO2} Berti, M.; {Bellini, E.; Bonvin, C.; Kunz, M.; Viel, M.; Zumalacarregui, M.}
	 Reconstructing the dark energy density in light of DESI BAO observations. \textit{Phys. Rev. D} \textbf{2025}, \textit{112}, 023518. 
	
	\bibitem{celiaBAO1} Escamilla-Rivera, C.;  Sandoval-Orozco, R. $f(T)$ gravity after DESI Baryon acoustic oscillation and DES supernovae 2024 data. \emph{J. High Energy Astrophys.} \textbf{2024}, \textit{42}, 217--221. 
	
	
	\bibitem{celiaBAO2} Aguilar, A.;  Escamilla-Rivera, C.;  Said, J.L.;  Mifsud, J. Non-fluid like Boltzmann code architecture for early times $f(T)$ cosmologies. {\emph{arXiv}} \textbf{2024}, arXiv:2403.13708. 
	
	
	\bibitem{celiaBAO3} Sandoval-Orozco, R.;  Escamilla-Rivera, C.; Briffa, R.;  Said, J.L. Testing $f(T)$ cosmologies with HII Hubble diagram and CMB distance priors. \emph{Phys. Dark Universe} \textbf{2024}, \textit{46}, 101641. 

		
	
\end{thebibliography}
\end{document}